\documentclass[final,3p,times]{elsarticle}
\usepackage{amssymb}
\usepackage{amsmath}
\usepackage{amsthm}
\usepackage{subfigure}
\usepackage{graphicx}

\biboptions{numbers,sort&compress}
\usepackage{hyperref}
\usepackage{booktabs}
\makeatletter

\newcommand{\Rmnum}[1]{\expandafter\@slowromancap\romannumeral #1@}
\makeatother

\usepackage{anysize}
\usepackage{setspace}
\marginsize{2.5cm}{2.2cm}{0.8cm}{1.2cm}
\journal{}

\begin{document}

\begin{frontmatter}



\title{Generalized Darboux transformation and higher-order rogue wave solutions of the coupled Hirota equations }

\author{Xin Wang}
\author{Yong Chen\corref{cor1}}
\ead{ychen@sei.ecnu.edu.cn}

\cortext[cor1]{Corresponding author.}

\address{Shanghai Key Laboratory of Trustworthy Computing, East China Normal University, Shanghai, 200062, People's Republic of China}

\begin{abstract}
This paper is dedicated to study higher-order rogue wave solutions of the coupled Hirota equations with high-order nonlinear effects
like the third dispersion, self-steepening and stimulated Raman scattering terms.
By using the generalized Darboux transformation, a unified representation of $N$th-order rogue wave solution
with 3$N$+1 free parameters is obtained. In particular, the first-order rogue wave solution containing
polynomials of fourth order, and the second-order rogue wave solution consisting of polynomials of eighth order
are explicitly presented. Through the numerical plots, we show that four or six fundamental rogue waves can
coexist in the second-order rogue waves. By adjusting the values of some free parameters, different kinds
of spatial-temporal distribution structures such as circular, quadrilateral, triangular, line and fundamental patterns
are exhibited. Moreover, we see that nine or twelve fundamental rogue waves can synchronously emerge in the third-order
rogue waves. The more intricate spatial-temporal distribution shapes are shown via adequate
choices of the free parameters. Several wave characteristics such as the amplitudes and the coordinate positions of the
highest peaks in the rogue waves are discussed.
\end{abstract}

\end{frontmatter}



\section{Introduction}  

Rogue waves (also known as freak waves and other similar names), initially termed as mysterious and exceptionally
disastrous oceanic surface waves, have been the
powerful and computational topic for more than a decade \cite{01,02,04,05,06,07}.
One of the key characteristic of the rogue waves is that their height or steepness can usually attain
two or three times greater than the average wave state, and they appear from nowhere and
disappear without a trace \cite{08}.
In addition to the deep ocean \cite{01,05}, rogue waves have also moved to other branches of physics,
such as optics and lasers \cite{04,09}, capillary waves \cite{10}, plasma physics \cite{11},
Bose-Einstein condensates \cite{12}, hydrodynamics \cite{13}, atmosphere \cite{14}, and so forth \cite{15}.

The simplest mathematical description of a rogue wave is the so-called Peregrine soliton, which is an
analytical rational solution of the nonlinear Schr\"{o}dinger (NLS) equation \cite{16}. In contrast to the
Akhmediev breathers (ABs) and Ma solitons \cite{17,18}, Peregrine soliton is localized in both space and time. It has
been experimentally observed in the water-wave tank \cite{13}, nonlinear optics fiber \cite{19}, and even in
noiselike-pulse laser emission \cite{20}.
Later, in the past years, a sequence of nonlinear models have been investigated to possess
lower or higher-order rogue wave solutions, the NLS equation \cite{21,22,23,24,25,26,28,30s,31,32},
derivative NLS equation \cite{33,34,chow},
Hirota equation \cite{36,37}, Sasa-Satsuma equation \cite{38},
Lakshmanan-Porsezian-Daniel equation \cite{41}, Gross-Pitaevskii equation \cite{42,43,44},
Davey-Stewartson equation \cite{48} and so on \cite{49,50}.
However, a complete understanding of the formation mechanism for the intricate rogue wave phenomena
is unclear, because of the difficult and hazardous observational conditions \cite{08}.

Recently, there has been noticeably surge of interest in studying rogue waves in coupled systems, since
many complex systems must comprise several waves with different modes or frequencies instead of a single one,
when considering the significant interaction processes \cite{42,51,52,53,lak,54,55,56,57,58,59,60}.
In fact, rogue waves in coupled systems can present
the more abundant dynamical structures than in the scalar ones, such as the dark rogue waves \cite{42},
the interactions between rogue waves and dark-bright solitons or breathers \cite{51,54},
the non-symmetric doubly localized rogue waves \cite{58},
the four-petaled rogue waves \cite{59}  and the
four-peaky shaped rogue waves \cite{60}.
Notably, very recently, the composite rogue waves in coupled systems
which can be well described  by the rational fourth-order or even higher-order polynomials solutions in mathematics
have attracted widespread attention \cite{52}.

In this paper, we consider the coupled Hirota (CH) equations in dimensional form, that is
\begin{align}
& \mathrm{i}u_{t}+\frac{1}{2}u_{xx}+(|u|^{2}+|v|^{2})u+\mathrm{i}\epsilon[u_{xxx}+(6|u|^{2}+3|v|^{2})u_{x}+3uv^{*}v_{x}]=0,
\label{01} \\
& \mathrm{i}v_{t}+\frac{1}{2}v_{xx}+(|u|^{2}+|v|^{2})v+\mathrm{i}\epsilon[v_{xxx}+(6|v|^{2}+3|u|^{2})v_{x}+3vu^{*}u_{x}]=0,
 \label{02}
\end{align}
where $u(x,t)$ and $v(x,t)$ are wave envelops, the asterisk represents complex conjugation,
$x$ is the transverse variable and $t$ is the propagation distance.
The real parameter $\epsilon$ represents the perturbation effects involving the third dispersion, self-steepening
and stimulated Raman scattering for the Manakov system. This system was originally formulated by Tasgal and
Potasek to describe the electromagnetic pulse propagation in coupled optical waveguides, and it is
more accurate than the Manakov system to describe the interaction process of two surface waves in the deep ocean,
and to describe the propagation of femtosecond optical pulses in the birefringent or two-mode nonlinear fibers \cite{61,62}.
Mathematically, Eqs. (1) and (2) are completely integrable in the sense of
their Lax pair \cite{61}, inverse scattering transformation \cite{63}, Painlev\'{e} analysis \cite{64},
Darboux transformation \cite{65}, Hirota bilinear form
and $N$-soliton solutions \cite{66}.
Recently, Chen presented the fundamental rogue wave, dark rogue wave and
composite rogue wave solutions \cite{62,67}, and we investigate the
interactional localized wave and rogue-wave pair
solutions of Eqs. (1) and (2) \cite{68,69}. However, to the best of our knowledge,
there are no reports on higher-order rogue wave solutions of the CH equations.

It is very known that higher-order rogue waves can be viewed as the nonlinear superposition or combination of
a fixed number of fundamental rogue waves. The experimental observations in a water tank and theoretical
classification of higher-order rogue waves for the scalar systems have been increasingly presented \cite{24,30s,37,70}.
Nonetheless, as is known to us, higher-order rogue waves in coupled systems have not been widely investigated.
Therefore, generalization to even higher-order rogue wave solutions of coupled systems is important and nontrivial.
The objective of this paper is to explore higher-order rogue wave
solutions of Eqs. (1) and (2) through the generalized Darboux transformation (DT) \cite{28,31,30s,33,60,68,71,72}.
By means of the generalized DT, a unified formula of $N$th-order rogue wave solution with 3$N$+1 free parameters is
derived by the direct iterative rule. Apart from the first-order rogue wave solution containing
polynomials of second or fourth order , we present that the second-order rogue wave solution can be composed by the
polynomials of eighth or twelfth order. Through the numerical plots, we show that four fundamental rogue waves
with quadrilateral, triangular, line and fundamental patterns,
as well as  six fundamental rogue waves with circular, two different types of quadrilateral,
triangular and fundamental patterns
can coexist in the second-order rogue waves. Moreover, the third-order rogue wave solution consists
of polynomials of 18th or 24th order. We exhibit that nine fundamental rogue waves with two different types of circular,
two different types of quadrilateral, triangular and fundamental patterns can emerge in the third-order rogue waves.
Further, twelve fundamental rogue waves with two different types of circular, two different types of pentagram,
two different types of quadrilateral,
triangular and fundamental patterns are explicitly shown in the third-order rogue waves.
In addition, by the numerical computation,
some wave characteristics such as the amplitudes and the coordinate positions of the
highest peaks in the rogue waves are discussed. Also, the perturbation influences produced by the small
real parameter $\epsilon$ on the higher-order rogue waves are demonstrated.

The arrangement of our paper is as follows. In section 2, we present the generalized DT and a unified
$N$th-order rogue wave solution of the CH equations. In section 3, some explicit rogue wave solutions and
numerical plots are shown. The conclusion is given in the final section.

\section{Generalized Darboux transformation}
In this section, according to the  Ablowitz-Kaup-Newell-Segur (AKNS) approach,
we begin with the following linear $3\times3$ matrix Lax pair
\begin{align}
&\Phi_{x}=U\Phi,\ \ U=\zeta U_{0}+U_{1}, \label{03} \\
&\Phi_{t}=V\Phi,\ \ V=\zeta^{3}V_{0}+\zeta^{2}V_{1}+\zeta V_{2}+V_{3},\label{04}
\end{align}
where
$$
U_{0}=\frac{1}{12\epsilon}\left(
\begin{array}{ccc}
-2\mathrm{i} & 0 & 0 \\
0 & \mathrm{i} & 0 \\
0 & 0 & \mathrm{i} \\
\end{array}
\right),\ \
U_{1}=
\left(
  \begin{array}{ccc}
    0 & -u & -v \\
    u^{*} & 0 & 0 \\
    v^{*} & 0 & 0 \\
  \end{array}
\right),\ V_{0}=\frac{1}{16\epsilon}U_{0},V_{1}=\frac{1}{8\epsilon}U_{0}+\frac{1}{16\epsilon}U_{1},
$$
$$
V_{2}=\frac{1}{4}
\left(
  \begin{array}{ccc}
    \mathrm{i}e & -\frac{u}{2\epsilon}-\mathrm{i}u_{x} &  -\frac{v}{2\epsilon}-\mathrm{i}v_{x}\\
    \frac{u^{*}}{2\epsilon}-\mathrm{i}u^{*}_{x} & -\mathrm{i}|u|^{2} & -\mathrm{i}vu^{*} \\
    \frac{v^{*}}{2\epsilon}-\mathrm{i}v^{*}_{x} & -\mathrm{i}uv^{*} & -\mathrm{i}|v|^{2} \\
  \end{array}
\right),\
V_{3}=
\left(
  \begin{array}{ccc}
    \epsilon(e_{1}+e_{2})+\frac{\mathrm{i}}{2}e & \epsilon e_{3}-\frac{\mathrm{i}}{2}u_{x} & \epsilon e_{4}-\frac{\mathrm{i}}{2}v_{x} \\
    -\epsilon e_{3}^{*}-\frac{\mathrm{i}}{2}u^{*}_{x} & -\epsilon e_{1}-\frac{\mathrm{i}}{2}|u|^{2} & \epsilon e_{5}-\frac{\mathrm{i}}{2}vu^{*} \\
    -\epsilon e_{4}^{*}-\frac{\mathrm{i}}{2}v^{*}_{x} & -\epsilon e^{*}_{5}-\frac{\mathrm{i}}{2}uv^{*} & -\epsilon e_{2}-\frac{\mathrm{i}}{2}|v|^{2} \\
  \end{array}
\right),
$$
with
$$
e=|u|^{2}+|v|^2,\  e_{1}=uu^{*}_{x}-u^{*}u_{x},\ e_{2}=vv^{*}_{x}-v^{*}v_{x},\ e_{3}=u_{xx}+2eu,\ e_{4}=v_{xx}+2ev,\ e_{5}=u^{*}v_{x}-vu^{*}_{x}.
$$
Here, $\Phi=(\psi(x,t),\phi(x,t),\chi(x,t))^{T}$ is the vector eigenfunction and $\zeta$ is the spectral parameter.
One can directly obtain Eqs. (1) and (2) from the compatibility condition of the above linear problem.

Let $\Phi_{1}=(\psi_{1},\phi_{1},\chi_{1})^{T}$ be a basic solution of the Lax pair equations (\ref{03}) and
(\ref{04}) at $u=u[0]$, $v=v[0]$ and $\zeta=\zeta_{1}$. Thus,
by resorting to the standard Darboux dressing procedure of the AKNS spectral problem \cite{62,73,74},
the Darboux transformation of the linear system (\ref{03}) and (\ref{04}) reads
\begin{align}
&\Phi[1]=T[1]\Phi,\ T[1]=(\zeta-\zeta_{1}^{*})I-(\zeta_{1}-\zeta_{1}^{*})X[0],\label{05}\\
&u[1]=u[0]+{\rm i}(\zeta_{1}-\zeta_{1}^{*})\frac{\psi_{1}[0]\phi_{1}[0]^{*}}{4\epsilon\Pi_{0}},\label{06}\\
&v[1]=v[0]+{\rm i}(\zeta_{1}-\zeta_{1}^{*})\frac{\psi_{1}[0]\chi_{1}[0]^{*}}{4\epsilon\Pi_{0}},\label{07}
\end{align}
where $(\psi_{1}[0],\phi_{1}[0],\chi_{1}[0])^{T}=(\psi_{1},\phi_{1},\chi_{1})^{T}=\Phi_{1}[0]$,
$$
X[0]=\frac{\Phi_{1}[0]\Phi_{1}[0]^{\dag}}{\Pi_{0}},\ \Pi_{0}=\Phi_{1}[0]^{\dag}\Phi_{1}[0]=|\psi_{1}[0]|^{2}+|\phi_{1}[0]|^{2}
+|\chi_{1}[0]|^{2}.
$$
Here $I$ is the $3\times3$ identity matrix, the dagger indicates complex conjugate transpose.

In the following, assume $\Phi_{l}=(\psi_{l},\phi_{l},\chi_{l})^{T}$ ($1\leq l\leq N$) be a basic solution
of the Lax pair equations (\ref{03}) and (\ref{04}) at $u=u[0]$, $v=v[0]$ and $\zeta=\zeta_{l}$. Repeating the
above process $N$ times, we end up at
\begin{align}
&\Phi[N]=T[N]T[N-1]\cdots T[1]\Phi,\ T[l]=(\zeta-\zeta_{l}^{*})I-(\zeta_{l}-\zeta_{l}^{*})X[l-1],\label{08}\\
&u[N]=u[0]+\sum_{l=1}^{N}{\rm i}(\zeta_{l}-\zeta_{l}^{*})\frac{\psi_{l}[l-1]\phi_{l}[l-1]^{*}}{4\epsilon\Pi_{l-1}},\label{09}\\
&v[N]=v[0]+\sum_{l=1}^{N}{\rm i}(\zeta_{l}-\zeta_{l}^{*})\frac{\psi_{l}[l-1]\chi_{l}[l-1]^{*}}{4\epsilon\Pi_{l-1}},\label{10}
\end{align}
where $(\psi_{l}[l-1],\phi_{l}[l-1],\chi_{l}[l-1])^{T}=\Phi_{l}[l-1]$,
$$
\Phi_{l}[l-1]=T_{l}[l-1]T_{l}[l-2]\cdots T_{l}[1]\Phi_{l},\ T_{l}[j]=T[j]|_{\zeta=\zeta_{l}},\ 1\leq j\leq l-1,
$$
and
$$
X[l-1]=\frac{\Phi_{l}[l-1]\Phi_{l}[l-1]^{\dag}}{\Pi_{l-1}},\ \Pi_{l-1}=\Phi_{l}[l-1]^{\dag}\Phi_{l}[l-1]=|\psi_{l}[l-1]|^{2}+|\phi_{l}[l-1]|^{2}
+|\chi_{l}[l-1]|^{2}.
$$

Hence, based on the above facts, the generalized DT can be constructed as follows.
Let
$\Phi_{1}=\Phi_{1}(\zeta_{1}+\delta)$ be a basic solution of the Lax pair equations (\ref{03}) and (\ref{04})
at $u=u[0]$, $v=v[0]$ and $\zeta=\zeta_{1}+\delta$. We suppose that $\Phi_{1}$ can be expanded as Taylor series
at $\delta=0$,
\begin{equation}\label{11}
\Phi_{1}=\Phi_{1}^{[0]}+\Phi_{1}^{[1]}\delta+\Phi_{1}^{[2]}\delta^{2}+\cdots+\Phi_{1}^{[N]}\delta^{N}+\cdots,
\end{equation}
where $\Phi_{1}^{[j]}=(\psi_{1}^{[j]},\phi_{1}^{[j]},\chi_{1}^{[j]})=\frac{\displaystyle1}{\displaystyle j!}
\frac{\displaystyle\partial^{j}\Phi_{1}}{\displaystyle\partial \delta^{j}}\mid_{\delta=0} (j=0,1,2,\cdots).$

(1) The first-step generalized DT

It is obvious that $\Phi_{1}^{[0]}$ is a special solution of the Lax pair equations (\ref{03}) and (\ref{04})
at $u=u[0]$, $v=v[0]$ and $\zeta=\zeta_{1}$, so the first-step generalized DT holds that
\begin{align}
&\Phi[1]=T[1]\Phi,\ T[1]=(\zeta-\zeta_{1}^{*})I-(\zeta_{1}-\zeta_{1}^{*})X[0],\label{12}\\
&u[1]=u[0]+{\rm i}(\zeta_{1}-\zeta_{1}^{*})\frac{\psi_{1}[0]\phi_{1}[0]^{*}}{4\epsilon\Pi_{0}},\label{13}\\
&v[1]=v[0]+{\rm i}(\zeta_{1}-\zeta_{1}^{*})\frac{\psi_{1}[0]\chi_{1}[0]^{*}}{4\epsilon\Pi_{0}},\label{14}
\end{align}
where $(\psi_{1}[0],\phi_{1}[0],\chi_{1}[0])^{T}=(\psi_{1}^{[0]},\phi_{1}^{[0]},\chi_{1}^{[0]})^{T}=\Phi_{1}[0]$,
$$
X[0]=\frac{\Phi_{1}[0]\Phi_{1}[0]^{\dag}}{\Pi_{0}},\ \Pi_{0}=\Phi_{1}[0]^{\dag}\Phi_{1}[0]=|\psi_{1}[0]|^{2}+|\phi_{1}[0]|^{2}
+|\chi_{1}[0]|^{2}.
$$

(2) The second-step generalized DT

As the next step, we take the following limit
$$\begin{array}{l}
\lim\limits_{\delta\rightarrow0}\frac{\displaystyle T[1]|_{\zeta=\zeta_{1}+\delta}\Phi_{1}}{\displaystyle\delta}=\lim\limits_{\delta\rightarrow0}
\frac{\displaystyle(\delta+T_{1}[1])\Phi_{1}}{\displaystyle\delta}=
\Phi_{1}^{[0]}+T_{1}[1]\Phi_{1}^{[1]}\equiv\Phi_{1}[1],
\end{array}$$
then the second-order generalized DT turns out that
\begin{align}
&\Phi[2]=T[2]T[1]\Phi,\ T[2]=(\zeta-\zeta_{1}^{*})I-(\zeta_{1}-\zeta_{1}^{*})X[1],\label{15}\\
&u[2]=u[1]+{\rm i}(\zeta_{1}-\zeta_{1}^{*})\frac{\psi_{1}[1]\phi_{1}[1]^{*}}{4\epsilon\Pi_{1}},\label{16}\\
&v[2]=v[1]+{\rm i}(\zeta_{1}-\zeta_{1}^{*})\frac{\psi_{1}[1]\chi_{1}[1]^{*}}{4\epsilon\Pi_{1}},\label{17}
\end{align}
where $(\psi_{1}[1],\phi_{1}[1],\chi_{1}[1])^{T}=\Phi_{1}[1]$,
$$
X[1]=\frac{\Phi_{1}[1]\Phi_{1}[1]^{\dag}}{\Pi_{1}},\ \Pi_{1}=\Phi_{1}[1]^{\dag}\Phi_{1}[1]=|\psi_{1}[1]|^{2}+|\phi_{1}[1]|^{2}
+|\chi_{1}[1]|^{2}.
$$

(3) The third-step generalized DT

Similarly, considering the following limit
$$\lim\limits_{\delta\rightarrow0}\frac{\displaystyle [T[2]T[1]]|_{\zeta=\zeta_{1}+\delta}\Phi_{1}}{\displaystyle\delta^{2}}=
\lim\limits_{\delta\rightarrow0}\frac{\displaystyle (\delta+T_{1}[2])(\delta+T_{1}[1])\Phi_{1}}{\displaystyle\delta^{2}}=
\Phi_{1}^{[0]}+(T_{1}[2]+T_{1}[1])\Phi_{1}^{[1]}+T_{1}[2]T_{1}[1]\Phi_{1}^{[2]}\equiv\Phi_{1}[2],
$$
the third-order generalized DT yields
\begin{align}
&\Phi[3]=T[3]T[2]T[1]\Phi,\ T[3]=(\zeta-\zeta_{1}^{*})I-(\zeta_{1}-\zeta_{1}^{*})X[2],\label{18}\\
&u[3]=u[2]+{\rm i}(\zeta_{1}-\zeta_{1}^{*})\frac{\psi_{1}[2]\phi_{1}[2]^{*}}{4\epsilon\Pi_{2}},\label{19}\\
&v[3]=v[2]+{\rm i}(\zeta_{1}-\zeta_{1}^{*})\frac{\psi_{1}[2]\chi_{1}[2]^{*}}{4\epsilon\Pi_{2}},\label{20}
\end{align}
where $(\psi_{1}[2],\phi_{1}[2],\chi_{1}[2])^{T}=\Phi_{1}[2]$,
$$
X[2]=\frac{\Phi_{1}[2]\Phi_{1}[2]^{\dag}}{\Pi_{2}},\ \Pi_{2}=\Phi_{1}[2]^{\dag}\Phi_{1}[2]=|\psi_{1}[2]|^{2}+|\phi_{1}[2]|^{2}
+|\chi_{1}[2]|^{2}.
$$

(4) The $N$th-step generalized DT

Taking all the above into account, and
proceeding in this way one by one, we have the general case, that is
$$
\Phi_{1}[l]=\Phi_{1}^{[0]}+\sum_{j=1}^{l}T_{1}[j]\Phi_{1}^{[1]}+\sum_{j=1}^{l}\sum_{k=1}^{j-1}T_{1}[j]T_{1}[k]\Phi_{1}^{[2]}
+\cdots+T_{1}[l]T_{1}[l-1]\cdots T_{1}[1]\Phi_{1}^{[l]},\ 1\leq l\leq N,
$$
\begin{align}
&\Phi[N]=T[N]T[N-1]\cdots T[1]\Phi,\ T[l]=(\zeta-\zeta_{1}^{*})I-(\zeta_{1}-\zeta_{1}^{*})X[l-1],\label{21}\\
&u[N]=u[0]+\sum_{l=1}^{N}{\rm i}(\zeta_{1}-\zeta_{1}^{*})\frac{\psi_{1}[l-1]\phi_{1}[l-1]^{*}}{4\epsilon\Pi_{l-1}},\label{22}\\
&v[N]=v[0]+\sum_{l=1}^{N}{\rm i}(\zeta_{1}-\zeta_{1}^{*})\frac{\psi_{1}[l-1]\chi_{1}[l-1]^{*}}{4\epsilon\Pi_{l-1}},\label{23}
\end{align}
where $(\psi_{1}[l-1],\phi_{1}[l-1],\chi_{1}[l-1])^{T}=\Phi_{1}[l-1]$,
$$
X[l-1]=\frac{\Phi_{1}[l-1]\Phi_{1}[l-1]^{\dag}}{\Pi_{l-1}},\ \Pi_{l-1}=\Phi_{1}[l-1]^{\dag}\Phi_{1}[l-1]=|\psi_{1}[l-1]|^{2}+|\phi_{1}[l-1]|^{2}
+|\chi_{1}[l-1]|^{2}.
$$

{\bf Remark 1.} It is notable to point out
that the expressions (\ref{22})-(\ref{23}) give rise to a unified $N$th-order rogue wave solution of Eqs. (1) and (2).
In the next section, we will present some concrete rogue wave solutions consisting of higher-order polynomials
to illustrate how to employ these formulas, and exhibit a series of figures to interpret the various
dynamical properties of the solutions.

\section{Rogue wave solutions}

As is known to all, rogue wave solutions are the limiting case of either ABs or Ma solitons
which can be generated from the plane waves. Thus, in this section, we start with a
plane-wave solution of Eqs. (1) and (2)
\begin{equation}\label{24}
u[0]=\exp[{\rm i}\theta_{1}],\ v[0]=\exp[{\rm i}\theta_{2}],
\end{equation}
where
$$
\theta_{1}=\frac{1}{2}x+\frac{(15-23\epsilon)}{8}t,\ \theta_{2}=-\frac{1}{2}x+\frac{(15+23\epsilon)}{8}t.
$$
After that, in order to seek out an adequate basic solution of the Lax pair equations (\ref{03}) and (\ref{04}),
we set $\zeta=6\sqrt{3}{\rm i}\epsilon(1+\theta^3)$, here $\theta$ is a small parameter.
Then under this
determined spectral parameter and the seed solution (\ref{24}), the basic solution matrix
of the Lax pair equations can be calculated as
\begin{equation}\label{25}
M(\theta)=D(M_{1},M_{2},M_{3}),
\end{equation}
where
$$
D={\rm diag}\left\{\exp[{\rm i}\displaystyle\frac{5}{4}t],\exp[{\rm i}(-\displaystyle\frac{1}{2}x-\frac{(5-23\epsilon)}{8}t)], \exp[{\rm i}(\displaystyle\frac{1}{2}x-\frac{(5+23\epsilon)}{8}t)]\right\},$$
$$
M_{i}=\left(
\begin{array}{c}
-\displaystyle\frac{1}{12\epsilon}[{\rm i}(\zeta-6\epsilon)-12\epsilon\xi_{i}][{\rm i}(\zeta+6\epsilon)-12\epsilon\xi_{i}]\exp[\omega_{i}]\\
\left[{\rm i}(\zeta-6\epsilon)-12\epsilon\xi_{i}\right]\exp[\omega_{i}]   \\
\left[{\rm i}(\zeta+6\epsilon)-12\epsilon\xi_{i}\right]\exp[\omega_{i}]   \\
\end{array}
\right),\ i=1,2,3,
$$
with
$$
\omega_{i}=\xi_{i}x+[\frac{1}{4}{\rm i}(\zeta+2)\xi_{i}^2+\frac{1}{24\epsilon}(\zeta^{2}+2\zeta-90\epsilon^{2})\xi_{i}+
\frac{1}{288\epsilon^{2}}{\rm i}(\zeta+2)(\zeta^{2}+108\epsilon^{2})]t,
$$
and $\xi_{i}$ satisfies a cubic equation
\begin{equation}\label{26}
\xi_{i}^{3}-(\frac{9}{4}\theta^{6}+\frac{9}{2}\theta^{3})\xi_{i}-\frac{3}{4}\sqrt{3}\theta^{9}-\frac{9}{4}\sqrt{3}\theta^{6}
-\frac{3}{2}\sqrt{3}\theta^{3}=0.
\end{equation}
Hereafter, let us define a basic solution of the Lax pair equations in the form of
\begin{equation}\label{27}
\Phi_{1}(\theta)=f\Phi_{11}+g\Phi_{12}+h\Phi_{13},
\end{equation}
where
$$\begin{array}{l}
f=f_{1}+f_{2}\theta^{3}+f_{3}\theta^{6}+\cdots+f_{N}\theta^{3N},\\
g=g_{1}+g_{2}\theta^{3}+g_{3}\theta^{6}+\cdots+g_{N}\theta^{3N},\\
h=h_{1}+h_{2}\theta^{3}+h_{3}\theta^{6}+\cdots+h_{N}\theta^{3N},
\end{array}
$$
$$\begin{array}{l}
\Phi_{11}=\displaystyle\frac{1}{18\epsilon}[M_{1}+M_{2}+M_{3}],\\
\Phi_{12}=\displaystyle\frac{2}{9\epsilon}\sqrt[3]{2}[\frac{M_{1}}{\theta}
+(-\frac{1}{2}-\frac{\sqrt{3}}{2}{\rm i})\frac{M_{2}}{\theta}+(-\frac{1}{2}+\frac{\sqrt{3}}{2}{\rm i})\frac{M_{3}}{\theta}],\\
\Phi_{13}=\displaystyle\frac{16}{9\epsilon}\sqrt[3]{4}[\frac{M_{1}}{\theta^{2}}
+(-\frac{1}{2}+\frac{\sqrt{3}}{2}{\rm i})\frac{M_{2}}{\theta^{2}}+(-\frac{1}{2}-\frac{\sqrt{3}}{2}{\rm i})\frac{M_{3}}{\theta^{2}}].
\end{array}$$
Here, $f_{i}$, $g_{i}$ and $h_{i}$ ($1\leq i\leq N$) are free parameters.
At this time, we prove that $\Phi_{1}(\theta)$ can be expanded as the Taylor series at $\theta=0$,
\begin{equation}\label{28}
\Phi_{1}(\theta)=\Phi_{1}^{[0]}+\Phi_{1}^{[1]}\theta^{3}+\Phi_{1}^{[2]}\theta^{6}+\cdots+\Phi_{1}^{[N]}\theta^{3N}+O(\theta^{3N}),
\end{equation}
where
$\Phi_{1}^{[j]}=(\psi_{1}^{[j]},\phi_{1}^{[j]},\chi_{1}^{[j]})=\frac{\displaystyle1}{\displaystyle 3j!}
\frac{\displaystyle\partial^{3j}\Phi_{1}}{\displaystyle\partial \theta^{3j}}\mid_{\theta=0} (j=0,1,2,\cdots)$. Here,
we give the explicit expressions of the first two terms coefficients, see appendix A.

It is straightforward to check that $\Phi_{1}^{[0]}$ is a nontrivial solution of the Lax pair
equations (\ref{03}) and (\ref{04}) at $u=u[0]$, $v=v[0]$ and $\zeta=\zeta_{1}=6\sqrt{3}{\rm i}\epsilon$.
So, by taking advantage of the formulas (\ref{13}) and (\ref{14}), the first-order rogue wave solution
can be calculated  with four free parameters in it, namely, $\epsilon$,
$f_{1}$, $g_{1}$ and $h_{1}$.
We now analyze the first-order rogue wave solution into two cases
based on the parameter $h_{1}$ chosen by zero or not.

{\bf Case 1}. $h_{1}=0$. In this case, we get the simple Peregrine soliton containing polynomials of second order.
By putting $g_{1}=1,f_{1}=0$, we arrive at
\begin{equation}\label{29}
u[1]=-\frac{(\sqrt{3}{\rm i}-1)}{2}\frac{(F_{1}+{\rm i}H_{1})}{D_{1}}\exp[{\rm i}\theta_{1}],\
v[1]=-\frac{(\sqrt{3}{\rm i}+1)}{2}\frac{G_{1}+{\rm i}K_{1}}{D_{1}}\exp[{\rm i}\theta_{2}],
\end{equation}
where
$$\begin{array}{l}
F_{1}=48x^{2}+64\sqrt{3}x+(3267\epsilon^{2}+36)t^{2}-(792\epsilon x+528\sqrt{3}\epsilon)t+32,\ H_{1}=48x-(396\epsilon+72)t+32\sqrt{3},\\
D_{1}=-48x^{2}-64\sqrt{3}x-(3267\epsilon^{2}+36)t^{2}+(792\epsilon x+528\sqrt{3}\epsilon)t-80,\\
G_{1}=-48x^{2}-64\sqrt{3}x-(3267\epsilon^{2}+36)t^{2}+(792\epsilon x+528\sqrt{3}\epsilon)t-32,\ K_{1}=48x+(-396\epsilon+72)t+32\sqrt{3}.
\end{array}
$$

{\bf Case 2}. $h_{1}\neq0$. At this point, it is found that the solution is made up of polynomials of fourth order.
By taking the free parameters such that
$h_{1}\neq0,f_{1}\neq0$ or $h_{1}\neq0,g_{1}\neq0$, we can work out
the solution which features a composite of two well-separated fundamental rogue waves.
For instance, let $h_{1}=1/100,f_{1}=10,g_{1}=0$, then it leads to
\begin{equation}\label{30}
\widetilde{u[1]}=-\frac{(\sqrt{3}{\rm i}-1)}{2}\frac{(\widetilde{F}_{1}+{\rm i}\widetilde{H}_{1})}{\widetilde{D}_{1}}\exp[{\rm i}\theta_{1}],\
\widetilde{v[1]}=-\frac{(\sqrt{3}{\rm i}+1)}{2}\frac{(\widetilde{G}_{1}+{\rm i}\widetilde{K}_{1})}{\widetilde{D}_{1}}\exp[{\rm i}\theta_{2}],
\end{equation}
where
$$\begin{array}{l}
\widetilde{F}_{1}=2304x^{4}+9216\sqrt{3}x^{3}+129024x^{2}+205312\sqrt{3}x+
81(363\epsilon^{2}+4)^{2}t^{4}+(-14256\epsilon(363\epsilon^{2}+4)x\\~~~~~-432\sqrt{3}\epsilon(14157\epsilon^{2}+196))t^{3}
+((940896\epsilon^{2}+3456)x^{2}+576\sqrt{3}(3663\epsilon^{2}+20)x+9528192\epsilon^{2}-57024\epsilon\\~~~~~-48384)t^{2}+
(-76032\epsilon x^{3}-241920\sqrt{3}\epsilon x^{2}+(-2211840\epsilon+6912)x-384\sqrt{3}(5251\epsilon-30))t+1242048,\\
\widetilde{H}_{1}=4608x^{3}+11520\sqrt{3}x^{2}+117504x+(864(33\epsilon+2)^{2}x-648(11\epsilon+2)(363\epsilon^{2}+4)t^{3}
+144\sqrt{3}(6237\epsilon^{2}\\~~~~~+780\epsilon+52))t^{2}+
(-(114048\epsilon+6912)x^{2}-1152\sqrt{3}(177\epsilon+14)x-990144\epsilon+114048)t+50048\sqrt{3},\\
\widetilde{D}_{1}=-2304x^{4}-9216\sqrt{3}x^{3}-135936x^{2}-214528\sqrt{3}x
-81(363\epsilon^{2}+4)^{2}t^{4}+
(14256\epsilon(363\epsilon^{2}+4)x\\~~~~~+432\sqrt{3}\epsilon(14157\epsilon^{2}+196))t^{3}+
(-(940896\epsilon^{2}+3456)x^{2}-576\sqrt{3}(3663\epsilon^{2}+20)x-9998640\epsilon^{2}+39744)t^{2}\\~~~~~+
(76032\epsilon x^{3}+241920\sqrt{3}\epsilon x^{2}+2325888\epsilon x+2085504\sqrt{3}\epsilon)t
-1206336,\\
\widetilde{G}_{1}=-2304x^{4}-9216\sqrt{3}x^{3}-129024x^{2}-205312\sqrt{3}x-81(363\epsilon^{2}+4)^{2}t^{4}
+(14256\epsilon(363\epsilon^{2}+4)x\\~~~~~+432\sqrt{3}\epsilon(14157\epsilon^{2}+196))t^{3}
+(-(940896\epsilon^{2}+3456)x^{2}
-576\sqrt{3}(3663\epsilon^{2}+20)x-9528192\epsilon^{2}-57024\epsilon\\~~~~~+48384)t^{2}
+(76032\epsilon x^{3}
+241920\sqrt{3}\epsilon x^{2}+(2211840\epsilon+6912)x+384\sqrt{3}(5251\epsilon+30))t-1242048,\\
\widetilde{K}_{1}=4608x^{3}+11520\sqrt{3}x^{2}+117504x+(864(33\epsilon-2)^{2}x
-648(11\epsilon-2)(363\epsilon^{2}+4)t^{3}
+144\sqrt{3}(6237\epsilon^{2}\\~~~~~-780\epsilon
+52))t^{2}
+((-114048\epsilon+6912)x^{2}-1152\sqrt{3}(177\epsilon-14)x-990144\epsilon-114048)t
+50048\sqrt{3}.\end{array}$$
Also, if we choose $h_{1}\neq0,f_{1}=0,g_{1}=0$, then the solution which is
characterized by a composite of two fundamental rogue waves which are closely intermingled with each other can be achieved
and here we refrain from presenting the explicit expression of it.

{\bf Remark 2.} Noteworthy, it should be pointed out that the validity of the above solutions (\ref{29}) and (\ref{30})
can be verified by putting them back into Eqs. (1) and (2), and
when we take the limit $\epsilon\rightarrow0$, they are reduced to the solutions of the Manakov system. Here,
we omit exhibiting the numerical plots of the aforementioned solutions which have been investigated by Chen and
us \cite{62,67,68,69}. In this paper, we focus on the higher-order rogue wave solutions of Eqs. (1) and (2), of which
the dynamic distribution structures will be much richer than those for the scalar ones.

At this time, performing the following limit
$$\begin{array}{l}
\lim\limits_{\theta\rightarrow0}\frac{\displaystyle T[1]|_{\zeta=6\sqrt{3}{\rm i}\epsilon(1+\theta^3)}\Phi_{1}}{\displaystyle\theta^{3}}=\lim\limits_{\theta\rightarrow0}
\frac{\displaystyle(6\sqrt{3}{\rm i}\epsilon\theta^{3}+T_{1}[1])\Phi_{1}}{\displaystyle\theta^{3}}=
6\sqrt{3}{\rm i}\epsilon\Phi_{1}^{[0]}+T_{1}[1]\Phi_{1}^{[1]}\equiv\Phi_{1}[1],
\end{array}$$
and in terms of the formulas (\ref{16}) and (\ref{17}), one obtains the second-order
rogue wave solution with seven free parameters in it, i.e.
$\epsilon$, $f_{1}$, $g_{1}$, $h_{1}$, $f_{2}$, $g_{2}$ and $h_{2}$. Likewise,
we classify the dynamic properties of the solution into two cases on
basis of the the parameter $h_{1}$ chosen by zero or not.

{\bf Case 1}. $h_{1}=0$. In this circumstance, the solution contains polynomials of eighth order and we
will show that four fundamental rogue waves can coexist in the second-order rogue waves.
Here, we consider four cases of the composite structures involving four fundamental rogue waves.

When we choose $g_{1}\neq0$, $f_{2}\neq0$ such as $g_{1}=1,f_{2}=10000$, then
the second-order rogue wave solution takes the form given below
\begin{equation}\label{31}
u[2]=\frac{(\sqrt{3}{\rm i}+1)}{2}\frac{(F_{2}+{\rm i}H_{2})}{D_{2}}\exp[{\rm i}\theta_{1}],\
v[2]=\frac{(\sqrt{3}{\rm i}-1)}{2}\frac{G_{2}+{\rm i}K_{2}}{D_{2}}\exp[{\rm i}\theta_{2}],
\end{equation}
where $F_{2},D_{2}$ and $G_{2}$ are three polynomials of eighth order in $x$ and $t$, as well as $H_{2}$ and $K_{2}$ are
two polynomials of seventh order in $x$ and $t$, see appendix B. By aid of the symbolic computation tool,
it is straightforward to check that (\ref{31}) agrees with Eqs. (1) and (2),  and
through the numerical
plots, we observe that there are four fundamental rogue waves arranging with a rhombus in
the second-order rogue waves, see Figs. 1(a) and 1(b). The amplitudes of the four highest peaks in $u$ component are
2.0806, 1.8712, 2.2361 and 1.8793, and occur at (6.5036,-0.0015),
(-1.6128,9.1116), (-3.0121,-8.9821) and (-10.0913,0.2432), respectively. For $v$ component,
the amplitudes of the four highest peaks are 2.0806, 2.2316, 1.8661 and 1.8793,
and arrive at (6.5040,0.0018), (-1.5025,8.9174), (-3.1587,-9.1815) and
(-10.1295,-0.2433), respectively.

\begin{figure}[!h]
\centering
\renewcommand{\figurename}{{\bf Fig.}}
{\includegraphics[height=4.5cm,width=16cm]{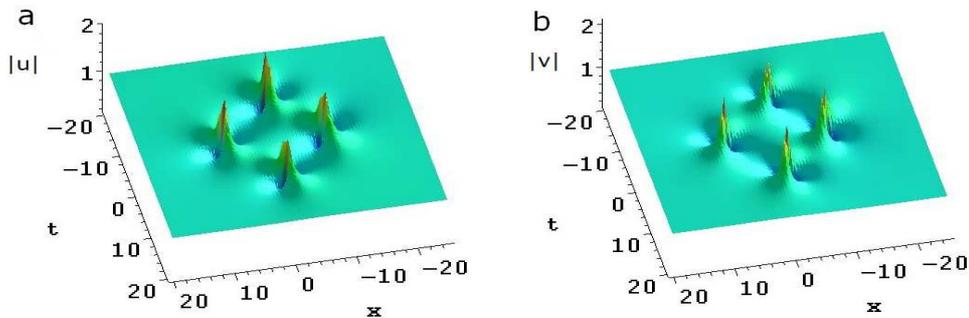}}
\caption{Evolution plots of the second-order rogue waves of quadrilateral pattern
in CH equations by choosing $\epsilon=0.01, f_{1}=0,g_{1}=1,h_{1}=0,
f_{2}=10000,g_{2}=0,h_{2}=0$.
(a) In $u$ component; (b) in $v$ component.}
\end{figure}

In the following, we refrain from writing down
the complicated expressions of the other second-order rogue wave solutions by choosing different values of the free parameters
and just show the interesting
dynamic structures, although it is not difficult to check the
validity of them with the aid of the symbolic computation tool.
Now we set $g_{1}\neq0,h_{2}\neq0$, then it is seen that in Figs. 2(a) and 2(b), two fundamental rogue waves
together with a composite rogue wave which is formed by the interaction of two fundamental ones
emerge with a triangular pattern on the spatial-temporal distribution;
when $g_{1}\neq0,f_{1}\neq0$, we see that
three rogue waves including a composite one arrange with a line pattern in Figs. 3(a) and 3(b); moreover,
let $g_{1}\neq0$ and the rest of the values be assumed to be zero, it is
displayed that four fundamental rogue waves intermingle with each other, see Figs. 4(a) and 4(b).
Particularly, it is calculated that the maximum amplitude of the peaks is achieved
at the composite rogue wave which is formed by the interaction of a fixed number of fundamental ones,
see table 1. We find that when the composite number is four, the maximum
amplitudes of the peaks in $u$ component and $v$ component
reach to 3.2697 and  3.2530, respectively,
which are greater than those of the situations with number two.

{\bf Remark 3.} It is worth mentioning that when by taking the limit $\epsilon\rightarrow0$, the second-order
rogue wave solution can also be reduced to that of the Manakov system. And because of the existence
of high-order nonlinear effects, the coordinate positions of some humps in the rogue waves
can be changed to a certain degree, particularly in the $x$ dimension. For instance,
by increasing the value of $\epsilon$, the effects can be observed more evidently, see Figs. 5(a) and 5(b).
The coordinate positions of the humps in $u$ component are (6.5029,-0.0003), (5.0775,8.8302),
(-10.1055,-9.3085) and (-10.0605,-0.2430), in $v$ component are (6.5071,0.0031), (5.0440,8.6565),
(-10.4401,-9.5337) and (-10.3018,-0.2437). Moreover, we  would also like to
note that when letting $g_{1}\neq0,g_{2}\neq0$, no new composite structures can be obtained
but the fundamental pattern and here we omit presenting it.

\begin{figure}[!h]
\centering
\renewcommand{\figurename}{{\bf Fig.}}
{\includegraphics[height=4.5cm,width=16cm]{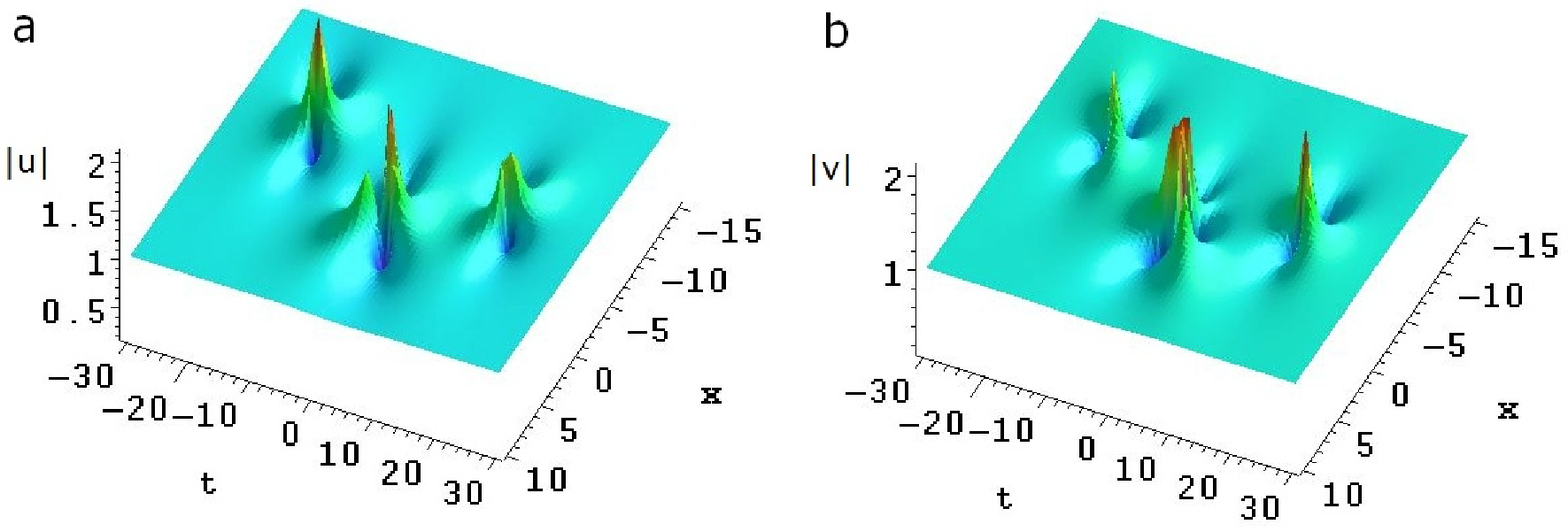}}
\caption{Evolution plots of the second-order rogue waves of triangular pattern
in CH equations by choosing $\epsilon=0.01, f_{1}=0,g_{1}=1,h_{1}=0,
f_{2}=0,g_{2}=0,h_{2}=10$.
(a) In $u$ component; (b) in $v$ component.}
\centering
\renewcommand{\figurename}{{\bf Fig.}}
{\includegraphics[height=4.5cm,width=16cm]{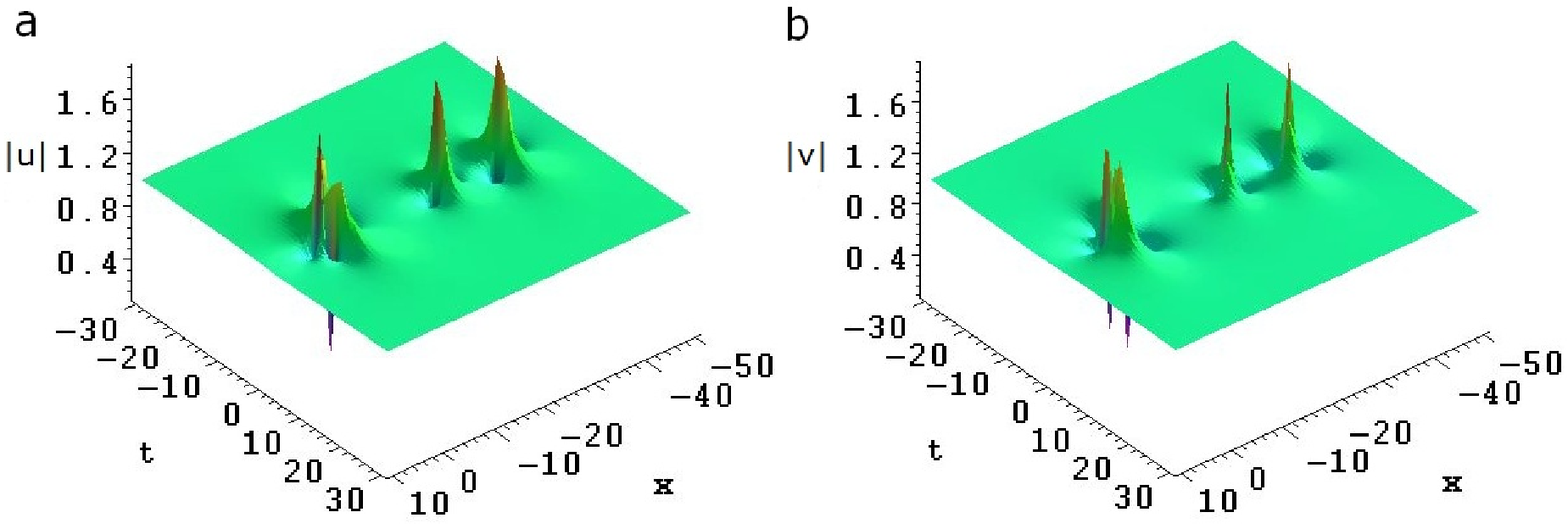}}
\caption{Evolution plots of the second-order rogue waves of line pattern
in CH equations by choosing $\epsilon=0.01, f_{1}=100,g_{1}=1,h_{1}=0,
f_{2}=0,g_{2}=0,h_{2}=0$.
(a) In $u$ component; (b) in $v$ component.}
\centering
\renewcommand{\figurename}{{\bf Fig.}}
{\includegraphics[height=4.5cm,width=16cm]{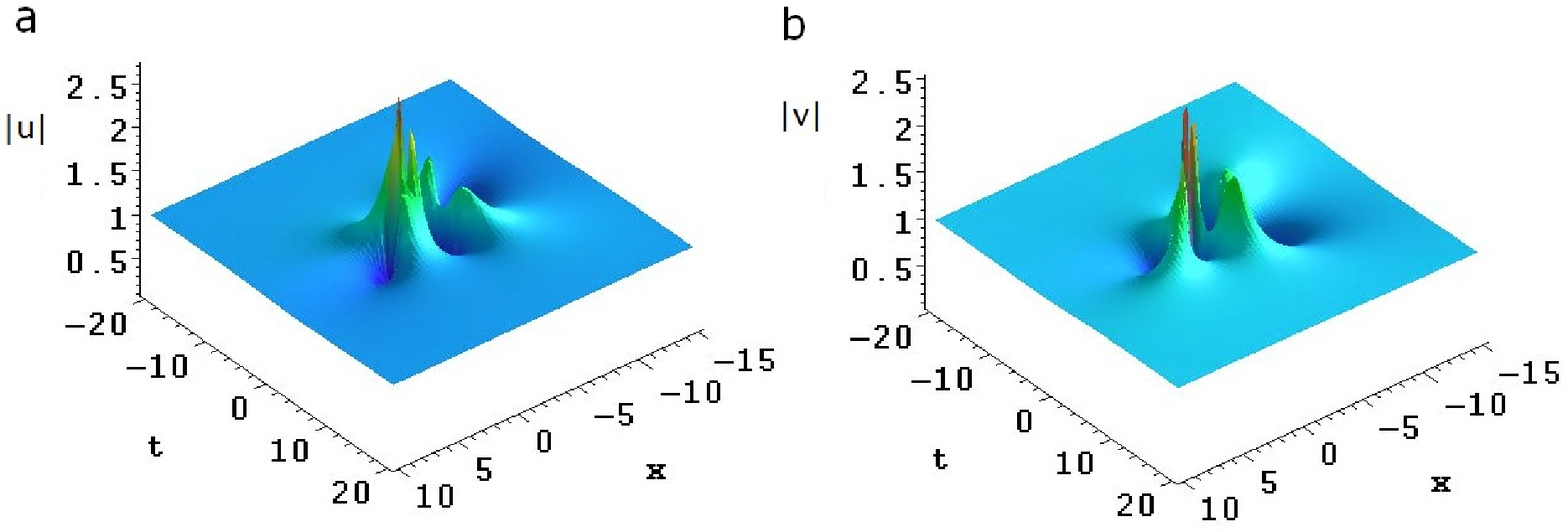}}
\caption{Evolution plots of the second-order rogue waves of fundamental pattern
in CH equations by choosing $\epsilon=0.01, f_{1}=0,g_{1}=1,h_{1}=0,
f_{2}=0,g_{2}=0,h_{2}=0$.
(a) In $u$ component; (b) in $v$ component.}
\end{figure}

\begin{table}[!h]
\begin{center}
\caption{Maximum amplitude of the peaks in the second-order rogue waves: four fundamental rogue waves case}
\begin{tabular}{lcll}
\toprule
pattern  &  composite no.   & ~~~~~~~~~~~~~~~~~~~~~$|u|$ & ~~~~~~~~~~~~~~~~~~~~~$|v|$  \\
\midrule
triangular & 2 &  $2.2643(x=-1.0537,t=-0.1127)$  & $2.2548(x=-1.0289,t=0.0985)$\\
line & 2  &  $2.8207,(x=-0.7985,t=0.0153)$ & $2.8209,(x=-0.8017,t=-0.0157)$ \\
fundamental & 4  &  $3.2697,(x=-0.7399,t=-0.0783)$ & $3.2530,(x=-0.7221,t=0.0753)$ \\
\bottomrule
\end{tabular}\\
\end{center}
\end{table}

\begin{figure}[!h]
\centering
\renewcommand{\figurename}{{\bf Fig.}}
{\includegraphics[height=4.5cm,width=16cm]{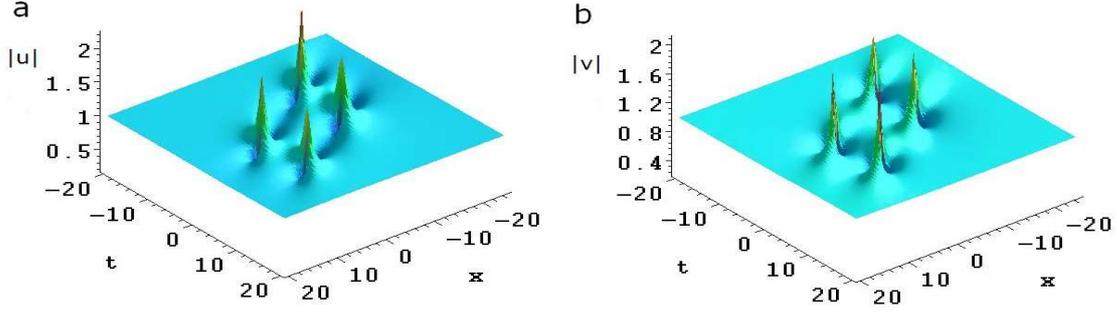}}
\caption{Evolution plots of the second-order rogue waves of quadrilateral pattern
in CH equations by choosing $\epsilon=0.1, f_{1}=0,g_{1}=1,h_{1}=0,
f_{2}=10000,g_{2}=0,h_{2}=0$.
(a) In $u$ component; (b) in $v$ component.}
\end{figure}

{\bf Case 2}. $h_{1}\neq0$. This time, the solution consists of polynomials of twelfth order and it will be
shown that six fundamental rogue waves can synchronously emerge in the second-order rogue waves.
We now discuss five cases of the composite structures involving fundamental rogue waves with number six.

By taking $h_{1}\neq0,f_{2}\neq0$, we observe that in Figs. 6(a) and 6(b),
one fundamental rogue wave is localized in the center, and five fundamental rogue waves are
distributed in the outer ring. The amplitude of the central hump in $u$ component is 2 and
occurs at (-1.1586,0.0093), in $v$ component is also 2 and arrive at (-1.1510,0.0064);
when setting $h_{1}\neq0,g_{2}\neq0$, the quadrilateral pattern \Rmnum{1} can be presented.
In Figs. 7(a) and 7(b), it is
exhibited that four fundamental rogue waves arrange with a rhombus
on the spatial-temporal distribution, and a composite rogue wave formed by the interaction of two
fundamental ones is localized in the interior of the rhombus;
when $h_{1}\neq0,f_{1}\neq0$,
in Figs. 8(a) and 8(b), the quadrilateral pattern \Rmnum{2}, namely,
the trapezium pattern is displayed;
when $h_{1}\neq0,g_{1}\neq0$, it is shown that a composite rogue wave constituted
by the interaction of four fundamental ones, as well as two fundamental rogue waves emerge with a
triangular pattern in Figs. 9(a) and 9(b); and when by taking $h_{1}\neq0$ and the other values equal to zero,
six fundamental rogue waves can merge with each other, see Figs. 10(a) and 10(b).
In addition, the maximum amplitudes of the peaks are given in table 2, and it is computed that
when six fundamental rogue waves mutually intermingle,
the maximum values can attain 3.7706 in $u$ component and 3.7732 in $v$ component, respectively.\\\\

\begin{figure}[!h]
\centering
\renewcommand{\figurename}{{\bf Fig.}}
{\includegraphics[height=4.5cm,width=16cm]{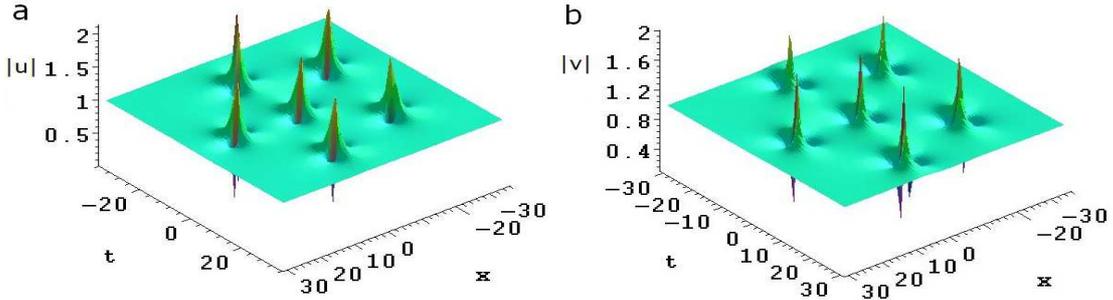}}
\caption{Evolution plots of the second-order rogue waves of circular pattern
in CH equations by choosing $\epsilon=0.01, f_{1}=0,g_{1}=0,h_{1}=0.01,
f_{2}=100000,g_{2}=0,h_{2}=0$.
(a) In $u$ component; (b) in $v$ component.}
\end{figure}

\begin{table}[!h]
\begin{center}
\caption{Maximum amplitude of the peaks in the second-order rogue waves: six fundamental rogue waves case}
\begin{tabular}{lcll}
\toprule
pattern  &  composite no.   & ~~~~~~~~~~~~~~~~~~~~~$|u|$ & ~~~~~~~~~~~~~~~~~~~~~$|v|$  \\
\midrule
quadrilateral \Rmnum{1} & 2 &  $2.2583,(x=-1.0564,t=-0.1114)$  & $2.2492,(x=-1.0318,t=0.0979)$\\
quadrilateral \Rmnum{2} & 2  &  $2.8313,(x=-0.7901,t=0.0161)$ & $2.8315,(x=-0.7936,t=-0.0165)$ \\
triangular & 4  &  $3.3219,(x=-0.7456,t=-0.0811)$ & $3.3017,(x=-0.7269,t=0.0782) $ \\
fundamental & 6  &  $3.7706,(x=-0.6098,t=-0.0014)$ & $3.7732,(x=-0.6104,t=-0.0061)$ \\
\bottomrule
\end{tabular}\\
\end{center}
\end{table}

\begin{figure}[!h]
\centering
\renewcommand{\figurename}{{\bf Fig.}}
{\includegraphics[height=4.5cm,width=16cm]{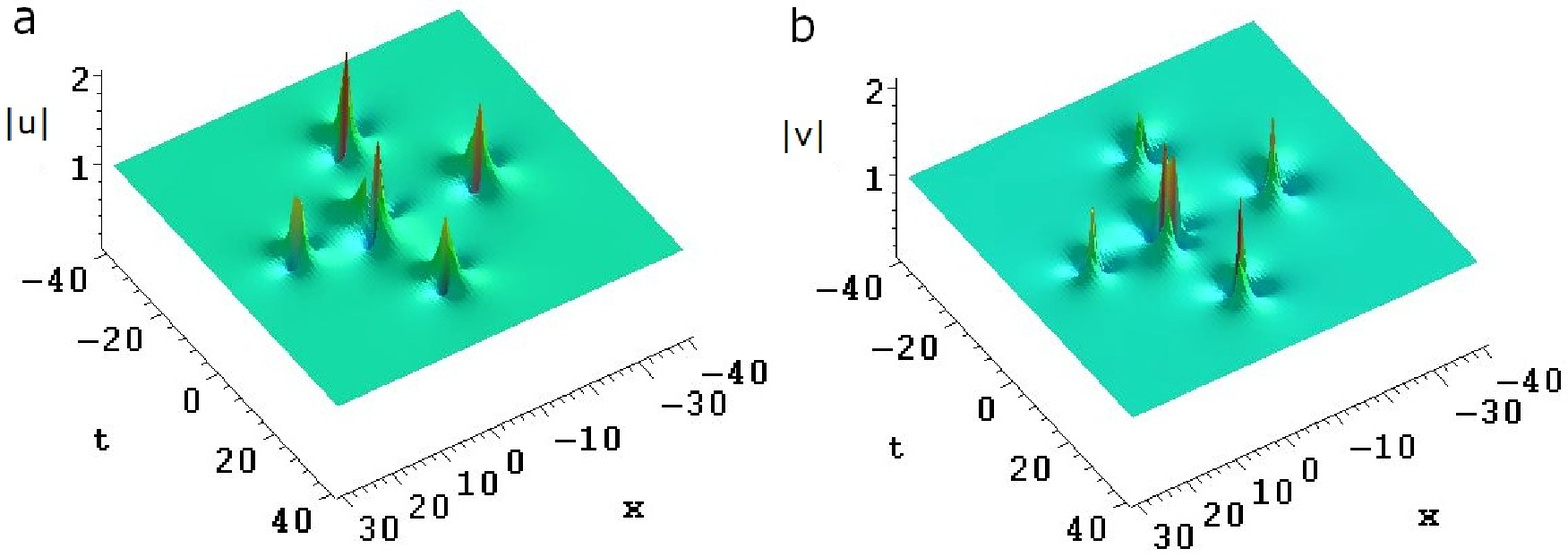}}
\caption{Evolution plots of the second-order rogue waves of quadrilateral pattern \Rmnum{1}
in CH equations by choosing $\epsilon=0.01, f_{1}=0,g_{1}=0,h_{1}=0.01,
f_{2}=0,g_{2}=1000,h_{2}=0$.
(a) In $u$ component; (b) in $v$ component.}
\centering
\renewcommand{\figurename}{{\bf Fig.}}
{\includegraphics[height=4.5cm,width=16cm]{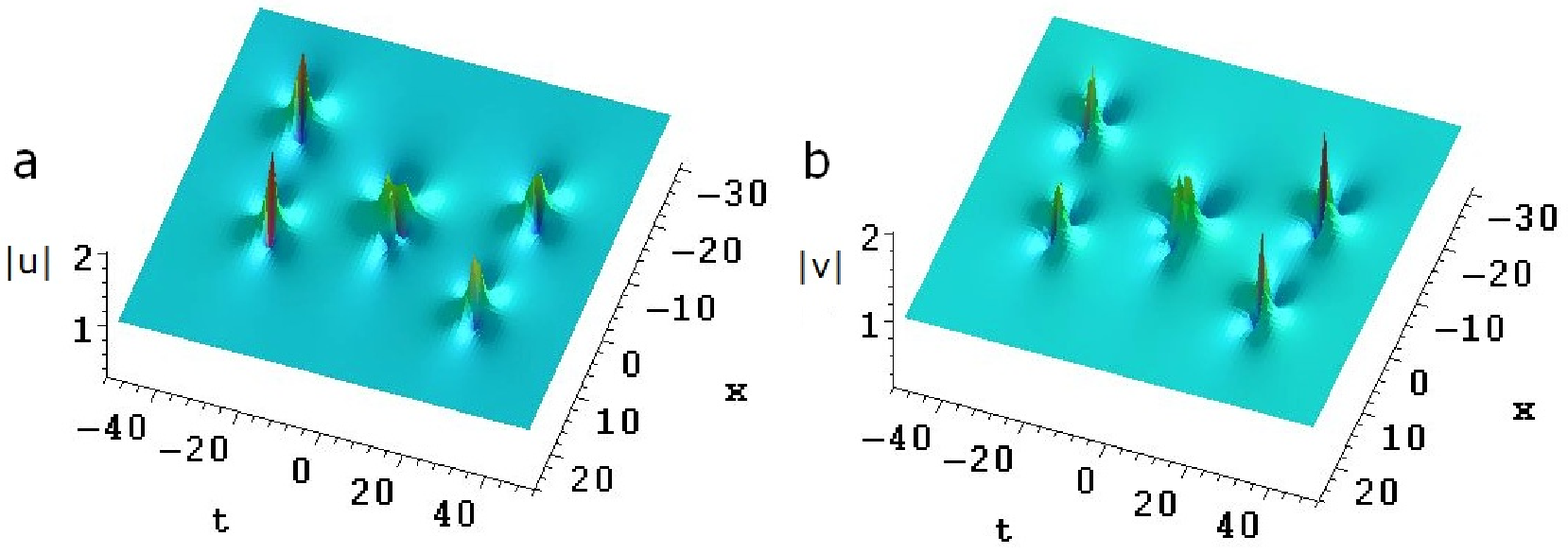}}
\caption{Evolution plots of the second-order rogue waves of quadrilateral pattern \Rmnum{2}
in CH equations by choosing $\epsilon=0.01, f_{1}=100,g_{1}=0,h_{1}=0.01,
f_{2}=0,g_{2}=0,h_{2}=0$.
(a) In $u$ component; (b) in $v$ component.}
\centering
\renewcommand{\figurename}{{\bf Fig.}}
{\includegraphics[height=4.5cm,width=16cm]{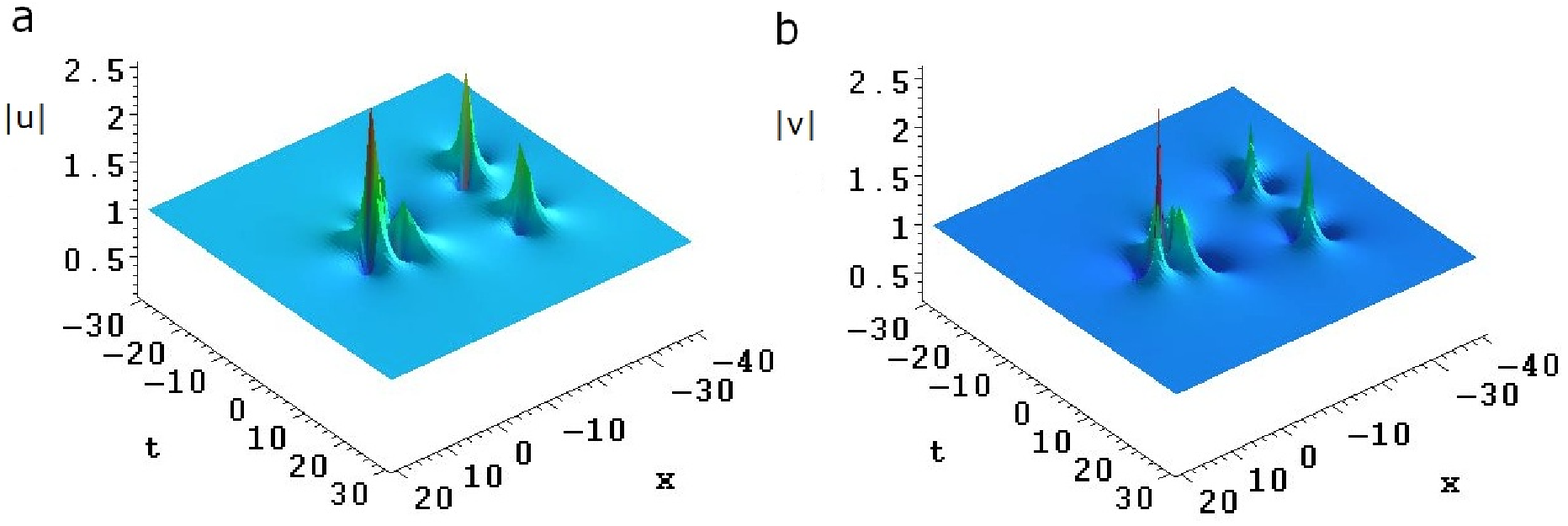}}
\caption{Evolution plots of the second-order rogue waves of triangular pattern
in CH equations by choosing $\epsilon=0.01, f_{1}=0,g_{1}=1,h_{1}=0.01,
f_{2}=0,g_{2}=0,h_{2}=0$.
(a) In $u$ component; (b) in $v$ component.}
\centering
\renewcommand{\figurename}{{\bf Fig.}}
{\includegraphics[height=4.5cm,width=16cm]{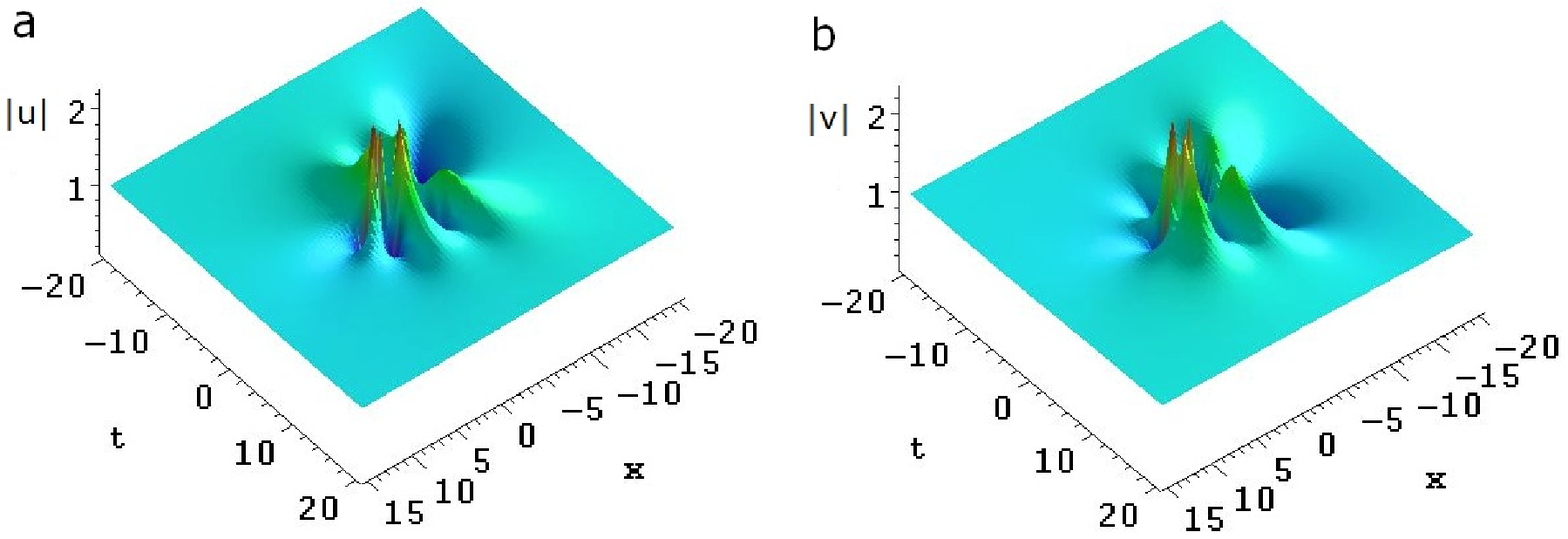}}
\caption{Evolution plots of the second-order rogue waves of fundamental pattern
in CH equations by choosing $\epsilon=0.01, f_{1}=0,g_{1}=0,h_{1}=0.01,
f_{2}=0,g_{2}=0,h_{2}=0$.
(a) In $u$ component; (b) in $v$ component.}
\end{figure}

By now, it should be emphasized that when
setting $h_{1}\neq0,h_{2}\neq0$, we also get the fundamental pattern without any new distribution
structures and here it is omitted. After that, we consider the limit such that
$$\begin{array}{r}
\lim\limits_{\theta\rightarrow0}\frac{\displaystyle [T[2]T[1]]|_{\zeta=6\sqrt{3}{\rm i}\epsilon(1+\theta^3)}\Phi_{1}}{\displaystyle\theta^{6}}=\lim\limits_{\theta\rightarrow0}
\frac{\displaystyle(6\sqrt{3}{\rm i}\epsilon\theta^{3}+T_{1}[2])(6\sqrt{3}{\rm i}\epsilon\theta^{3}+T_{1}[1])\Phi_{1}}{\displaystyle\theta^{6}}=
[6\sqrt{3}{\rm i}\epsilon]^{2}\Phi_{1}^{[0]}\\+6\sqrt{3}{\rm i}\epsilon(T_{1}[2]+T_{1}[1])\Phi_{1}^{[1]}+
T_{1}[2]T_{1}[1]\Phi_{1}^{[2]}  \equiv\Phi_{1}[2],~~~~~~~~~~~~
\end{array}$$
combining with the formulas (\ref{19}) and (\ref{20}), the third-order rogue wave solution with ten free parameters can be
obtained. Here, we refrain from presenting the expressions of $\Phi_{1}^{[2]}$ and the cumbersome solution,
although it is not difficult to check the validity of the solution
by aid of the symbolic computation tool. In a similar way, we classify
the third-order rogue wave solution into two cases through the parameter $h_{1}$ chosen by zero or not.

{\bf Case 1}. $h_{1}=0$. At this time, the solution is made up of polynomials of 18th order, and it
will be exhibited that nine fundamental rogue waves can coexist in the third-order rogue waves.
Next, we consider six cases of the composite structures involving nine fundamental rogue waves.

By setting $g_{1}\neq0,f_{3}\neq0$, the circular pattern \Rmnum{1} can be presented. We see that in Figs. 11(a) and
11(b), seven fundamental rogue waves are distributed in the outer ring, and a composite rogue wave
constituted by the interaction of two fundamental ones is localized in the center;
by taking $g_{1}\neq0,h_{3}\neq0$, the circular pattern \Rmnum{2} is obtained.
It is seen that five fundamental rogue waves in the outer ring,
together with a composite rogue wave in the center formed by the interaction of
four fundamental ones emerge on the spatial-temporal distribution, see Figs. 12(a) and 12(b);
when $g_{1}\neq0,f_{2}\neq0$, for the quadrilateral pattern \Rmnum{1},
we observe that nine well-separated fundamental rogue waves arrange with a rhombus, see Figs. 13(a) and 13(b);
when $g_{1}\neq0,h_{2}\neq0$, for the quadrilateral pattern \Rmnum{2}, it is shown that
four fundamental rogue waves arrange with a trapezium, and a composite rogue wave constituted by the
interaction of five fundamental ones is localized in the interior of the trapezium, see Figs. 14(a) and 14(b);
when $g_{1}\neq0,f_{1}\neq0$, the triangular pattern can be presented, see Figs. 15(a) and 15(b), the composite
rogue wave is formed by the interaction of six fundamental ones; when $g_{1}\neq0$ and the rest of the values
are set to be zero, then we observe that in Figs. 16(a) and 16(b), nine fundamental rogue waves
intermingle with each other, and the maximum amplitude in $u$ component reaches to 4.1566,
in $v$ component 4.1140. While when choosing $g_{1}\neq0,g_{2}\neq0$ or $g_{1}\neq0,g_{3}\neq0$,
there are no new composite structures but the fundamental pattern and here we omit presenting it.
Equally, the detailed numerical values of the
maximum amplitudes, composite numbers and coordinate positions are given,  see table 3.

\begin{figure}[!h]
\centering
\renewcommand{\figurename}{{\bf Fig.}}
{\includegraphics[height=4.5cm,width=16cm]{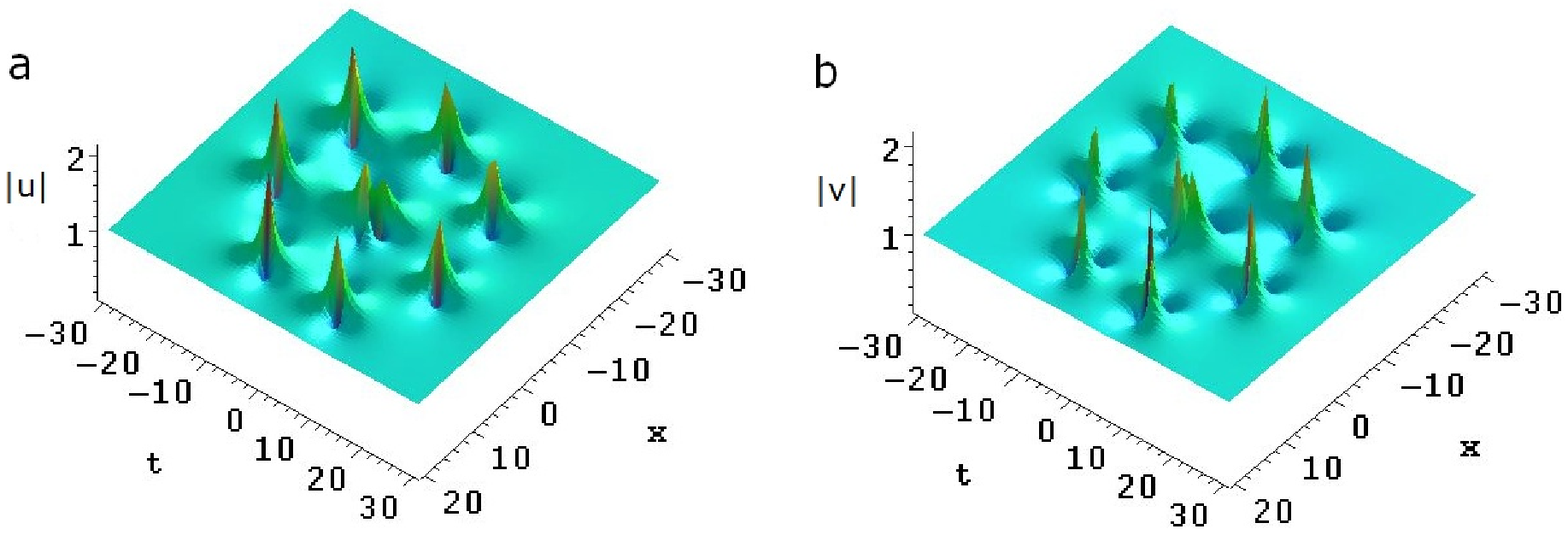}}
\caption{Evolution plots of the third-order rogue waves of circular pattern \Rmnum{1}
in CH equations by choosing $\epsilon=0.01, f_{1}=0,g_{1}=1,h_{1}=0,
f_{2}=0,g_{2}=0,h_{2}=0,f_{3}=10000000,g_{3}=0,h_{3}=0$.
(a) In $u$ component; (b) in $v$ component.}
\centering
\renewcommand{\figurename}{{\bf Fig.}}
{\includegraphics[height=4.5cm,width=16cm]{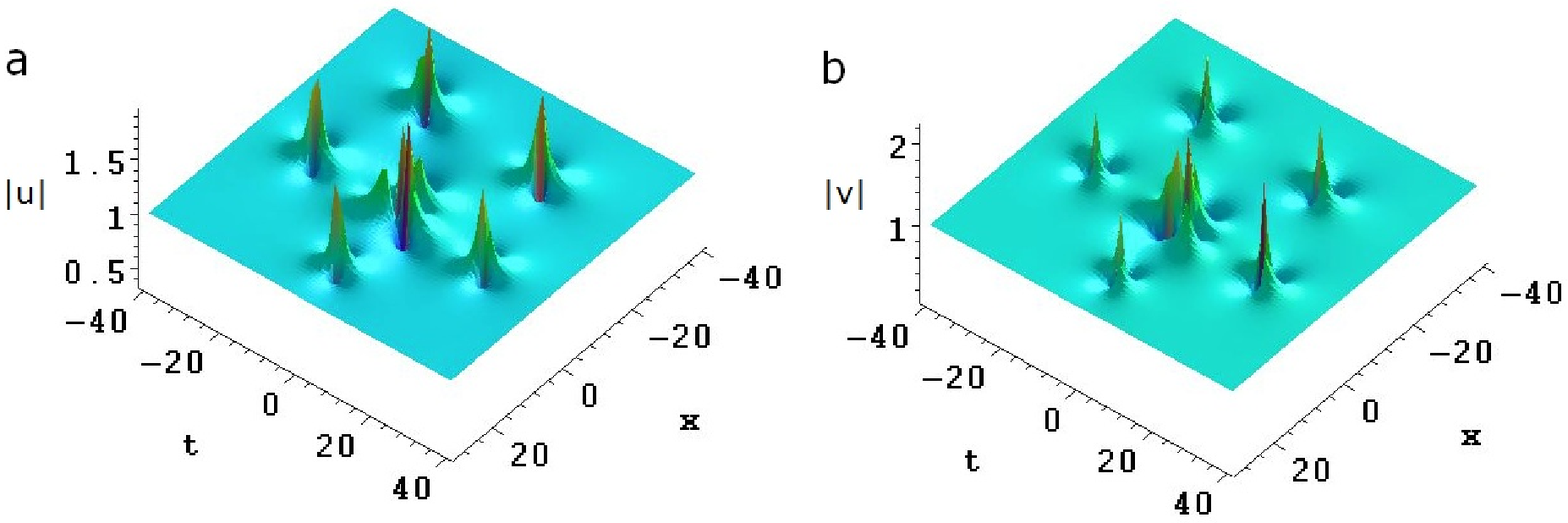}}
\caption{Evolution plots of the second-order rogue waves of circular pattern \Rmnum{2}
in CH equations by choosing $\epsilon=0.01, f_{1}=0,g_{1}=1,h_{1}=0,
f_{2}=0,g_{2}=0,h_{2}=0,f_{3}=0,g_{3}=0,h_{3}=10000$.
(a) In $u$ component; (b) in $v$ component.}
\end{figure}

\begin{figure}[!h]
\centering
\renewcommand{\figurename}{{\bf Fig.}}
{\includegraphics[height=4.5cm,width=16cm]{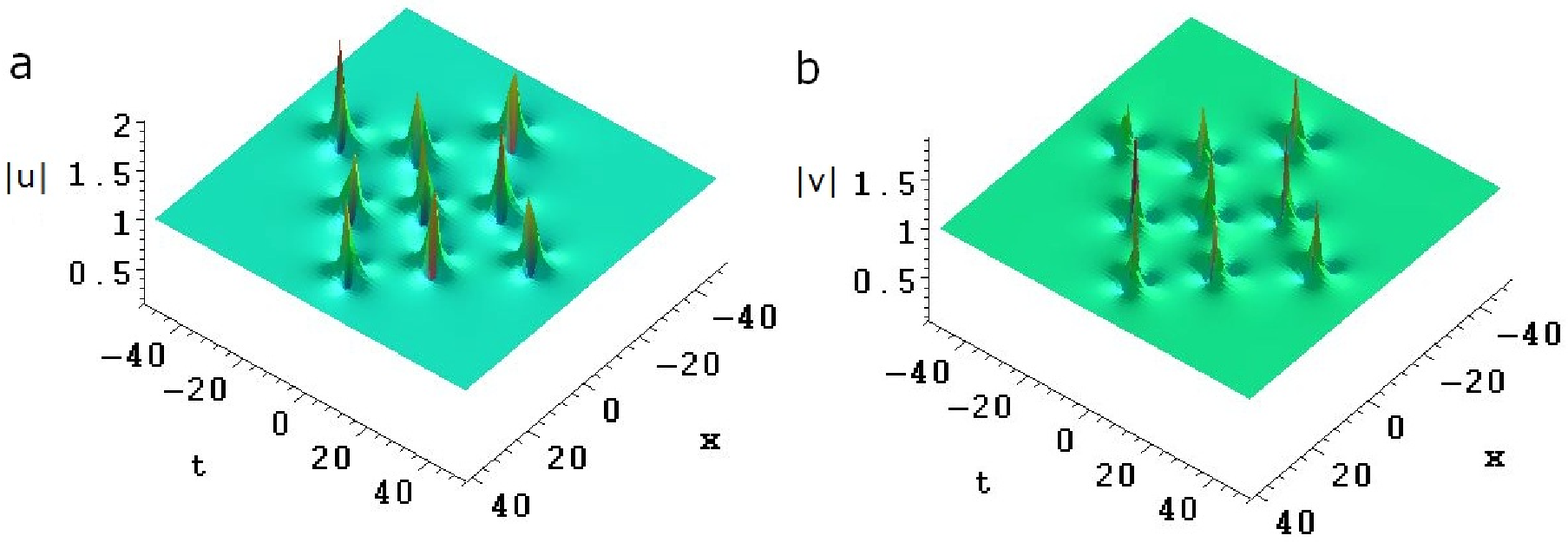}}
\caption{Evolution plots of the third-order rogue waves of quadrilateral pattern \Rmnum{1}
in CH equations by choosing $\epsilon=0.01, f_{1}=0,g_{1}=1,h_{1}=0,
f_{2}=100000,g_{2}=0,h_{2}=0,f_{3}=0,g_{3}=0,h_{3}=0$.
(a) In $u$ component; (b) in $v$ component.}
\centering
\renewcommand{\figurename}{{\bf Fig.}}
{\includegraphics[height=4.5cm,width=16cm]{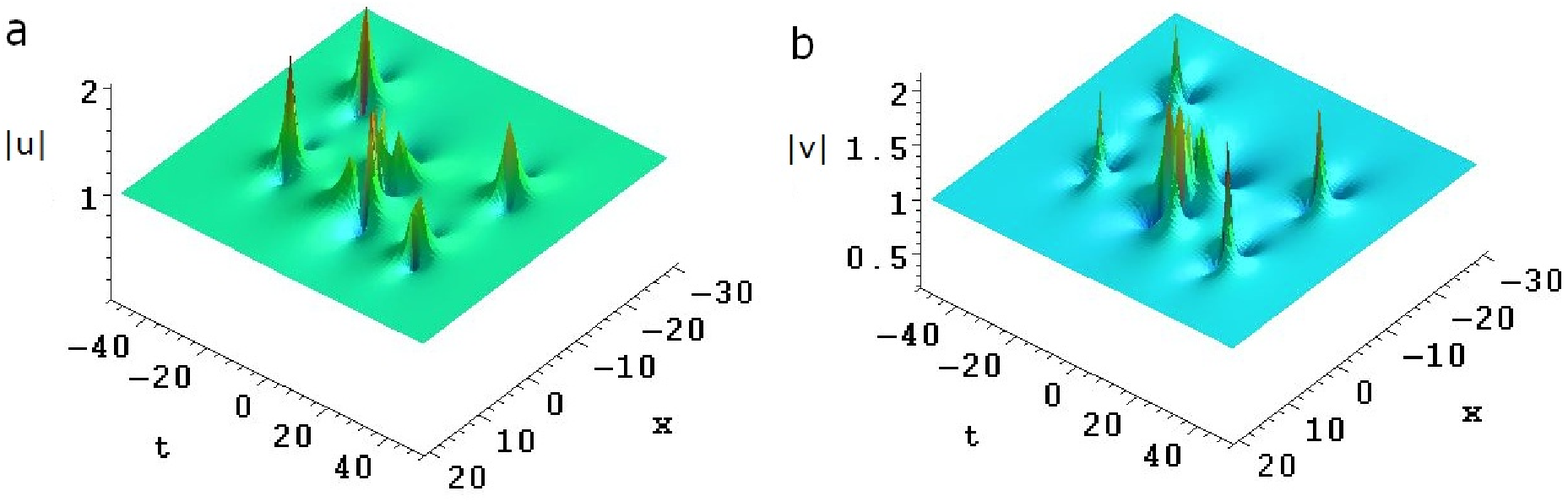}}
\caption{Evolution plots of the third-order rogue waves of quadrilateral pattern \Rmnum{2}
in CH equations by choosing $\epsilon=0.01, f_{1}=0,g_{1}=1,h_{1}=0,
f_{2}=0,g_{2}=0,h_{2}=10,f_{3}=0,g_{3}=0,h_{3}=0$.
(a) In $u$ component; (b) in $v$ component.}
\centering
\renewcommand{\figurename}{{\bf Fig.}}
{\includegraphics[height=4.5cm,width=16cm]{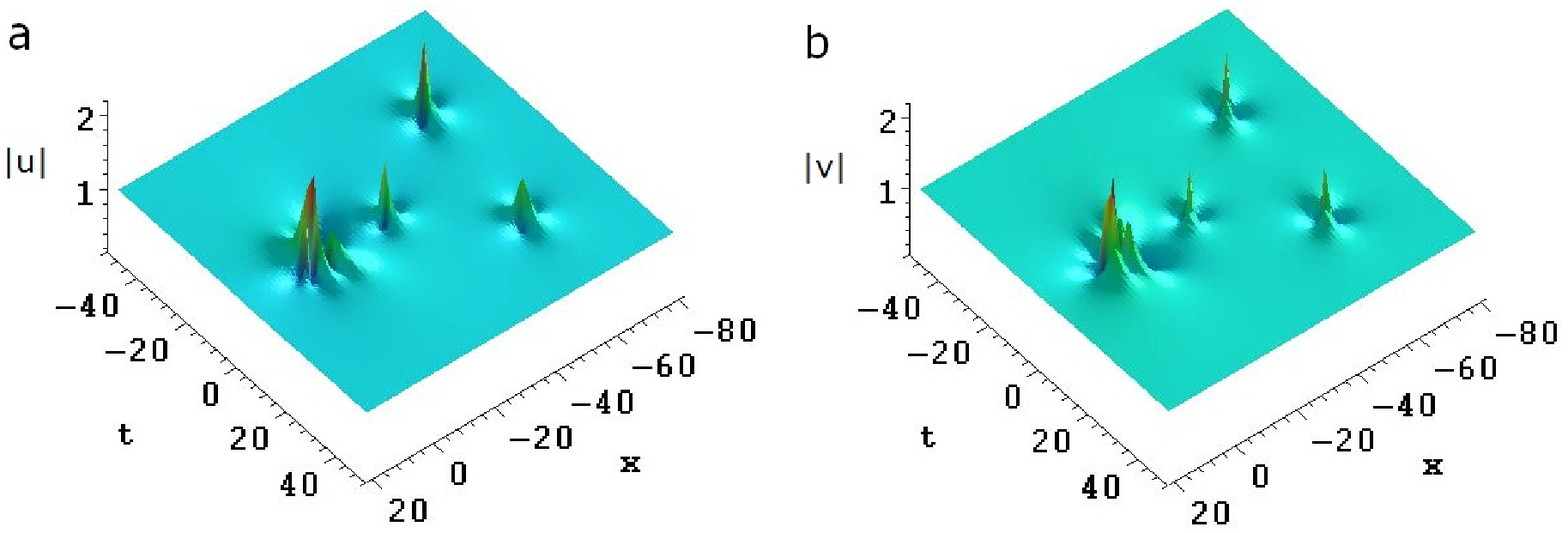}}
\caption{Evolution plots of the third-order rogue waves of triangular pattern
in CH equations by choosing $\epsilon=0.01, f_{1}=100,g_{1}=1,h_{1}=0,
f_{2}=0,g_{2}=0,h_{2}=0,f_{3}=0,g_{3}=0,h_{3}=0$.
(a) In $u$ component; (b) in $v$ component.}
\centering
\renewcommand{\figurename}{{\bf Fig.}}
{\includegraphics[height=4.5cm,width=16cm]{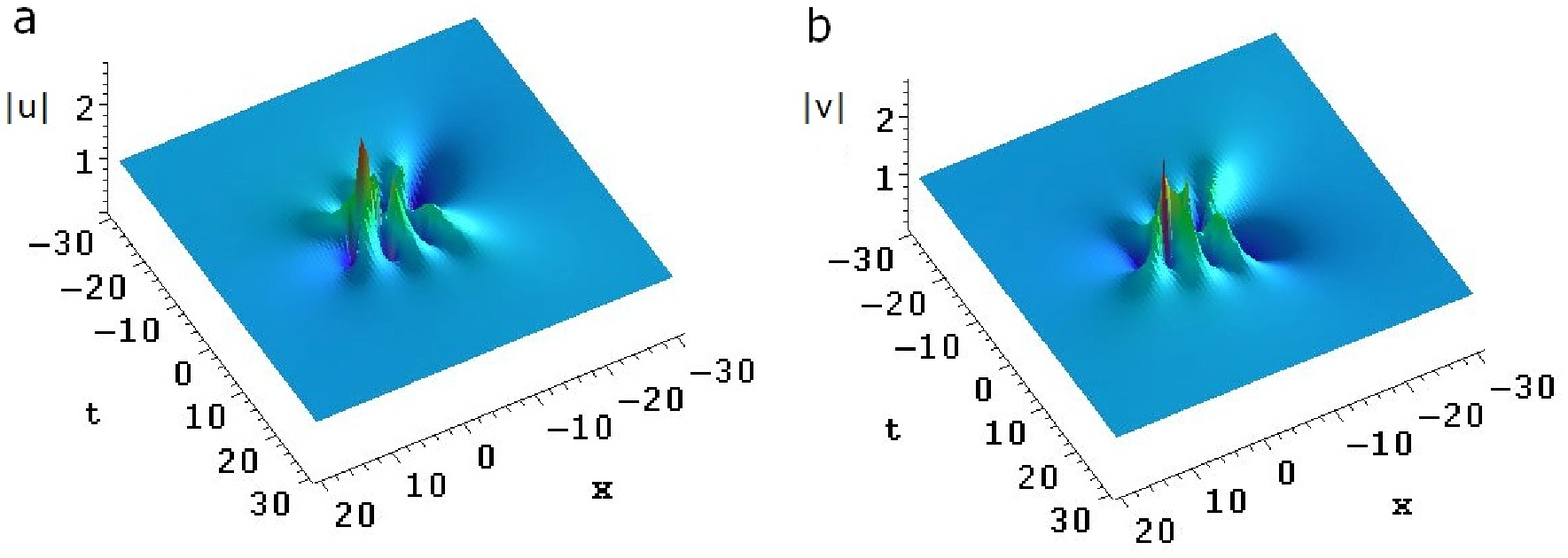}}
\caption{Evolution plots of the third-order rogue waves of fundamental pattern
in CH equations by choosing $\epsilon=0.01, f_{1}=0,g_{1}=1,h_{1}=0,
f_{2}=0,g_{2}=0,h_{2}=0,f_{3}=0,g_{3}=0,h_{3}=0$.
(a) In $u$ component; (b) in $v$ component.}
\end{figure}

\begin{table}[!h]
\begin{center}
\caption{Maximum amplitude of the peaks in the third-order rogue waves: nine fundamental rogue waves case}
\begin{tabular}{lcll}
\toprule
pattern  &  composite no.   & ~~~~~~~~~~~~~~~~~~~~~$|u|$ & ~~~~~~~~~~~~~~~~~~~~~$|v|$  \\
\midrule
circular \Rmnum{1} & 2 &  $2.8270,(x=-0.7905,t=0.0159)$  & $2.8272,(x=-0.7938,t=-0.0163)$\\
circular \Rmnum{2} & 4  &  $3.0403,(x=-0.8105,t=0.0077)$ & $3.0460,(x=-0.8192,t=-0.0201) $ \\
quadrilateral \Rmnum{2} & 5  &  $3.5877,(x=-0.6798,t=0.0532)$ & $3.6110,(x=-0.6949,t=-0.0566)$ \\
triangular & 6  &  $3.7527,(x=-0.6153,t=-0.0025)$ & $3.7542,(x=-0.6157,t=-0.0052) $ \\
fundamental & 9  &  $4.1566,(x=-0.5885,t=-0.0344)$ & $4.1140,(x=-0.5787,t=0.0289)$ \\
\bottomrule
\end{tabular}\\
\end{center}
\end{table}

{\bf Case 2}. $h_{1}\neq0$. At this moment, we obtain the solution containing polynomials of 24th order.
We will show that twelve fundamental rogue waves can synchronously emerge
in the third-order rogue waves, and eight cases of the composite structures involving twelve fundamental rogue waves
are explicitly shown.

Now by setting $h_{1}\neq0,f_{3}\neq0$, the circular pattern \Rmnum{1} is presented. It is seen that
in Figs. 17(a) and 17(b), a composite rogue wave formed by the interaction of four fundamental
ones is localized in the center, with eight fundamental rogue waves distributed in the
outer ring; by taking $h_{1}\neq0,g_{3}\neq0$, we show that the circular pattern \Rmnum{2}
is constituted by a composite rogue wave which is formed by the interaction of five fundamental
ones in the center, and seven fundamental rogue waves in the outer ring, see Figs. 18(a) and 18(b);
when $h_{1}\neq0,f_{2}\neq0$, eleven rogue waves including a composite one
arrange with a pentagram on the spatial-temporal distribution, see Figs. 19(a) and 19(b);
when $h_{1}\neq0,f_{2}\neq0,g_{1}\neq0$, we exhibit that twelve fundamental rogue waves
are nicely separated and emerge with a pentagram, see Figs. 20(a) and 20(b);
by taking $h_{1}\neq0,g_{2}\neq0$, the quadrilateral pattern \Rmnum{1} is presented. It is seen that
nine rogue waves including a composite one which is formed by the interaction
of four fundamental rogue waves arrange with a rhombus in Figs. 21(a) and 21(b);
while by letting $h_{1}\neq0,f_{1}\neq0$, the quadrilateral pattern \Rmnum{2}, namely, the
trapezium pattern is presented, see Figs. 22(a) and 22(b). The composite rogue wave in the middle
is constituted by the interaction of six fundamental ones; when choosing
$h_{1}\neq0,g_{1}\neq0$, a composite rogue wave together with three fundamental ones emerge with
a triangular pattern in Figs. 23(a) and 23(b); when $h_{1}\neq0$ and the rest of the values
are set to be zero, twelve fundamental rogue waves can merge with each other, see Figs. 24(a) and 24(b).
This time, the maximum amplitude in $u$ component attains 4.8530 and in $v$ component 4.8558, respectively.
While by choosing $h_{1}\neq0,h_{2}\neq0$ or $h_{1}\neq0,h_{3}\neq0$, the situation is trivial but the
fundamental pattern and we refrain from presenting it. Likewise,
the detailed numerical values of the corresponding index are shown, see table 4.

\begin{figure}[!h]
\centering
\renewcommand{\figurename}{{\bf Fig.}}
{\includegraphics[height=4.5cm,width=16cm]{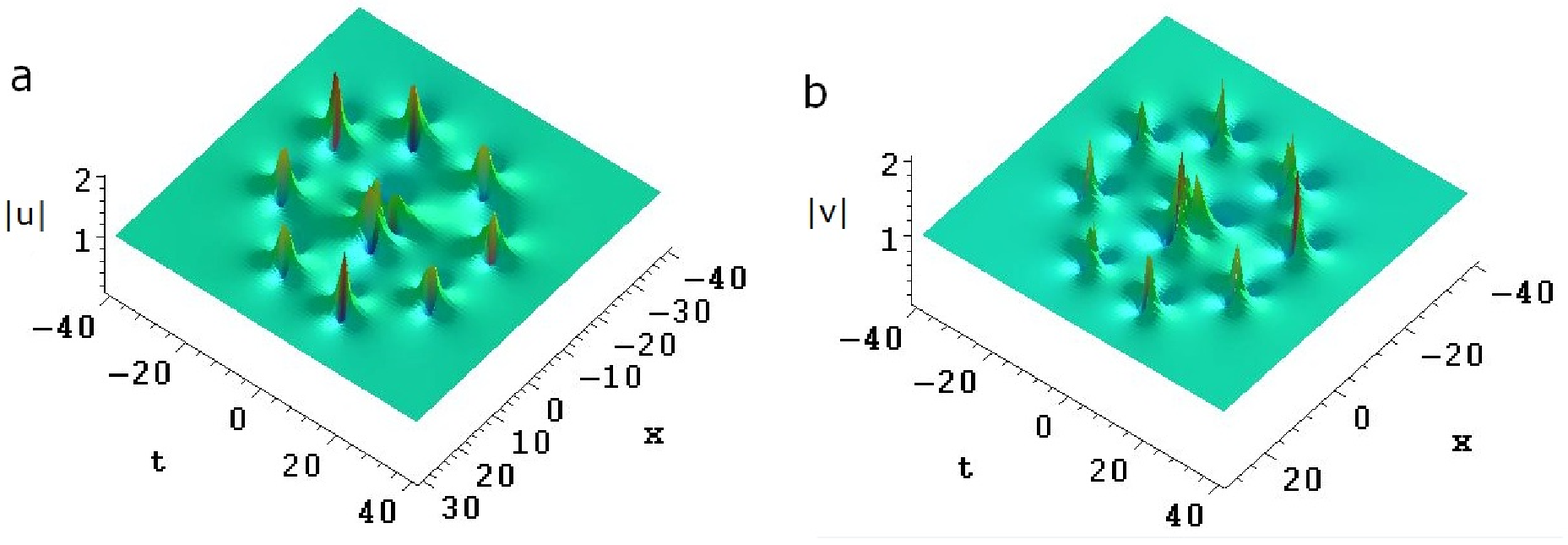}}
\caption{Evolution plots of the third-order rogue waves of circular pattern \Rmnum{1}
in CH equations by choosing $\epsilon=0.01, f_{1}=0,g_{1}=0,h_{1}=0.01,
f_{2}=0,g_{2}=0,h_{2}=0,f_{3}=10000000,g_{3}=0,h_{3}=0$.
(a) In $u$ component; (b) in $v$ component.}
\centering
\renewcommand{\figurename}{{\bf Fig.}}
{\includegraphics[height=4.5cm,width=16cm]{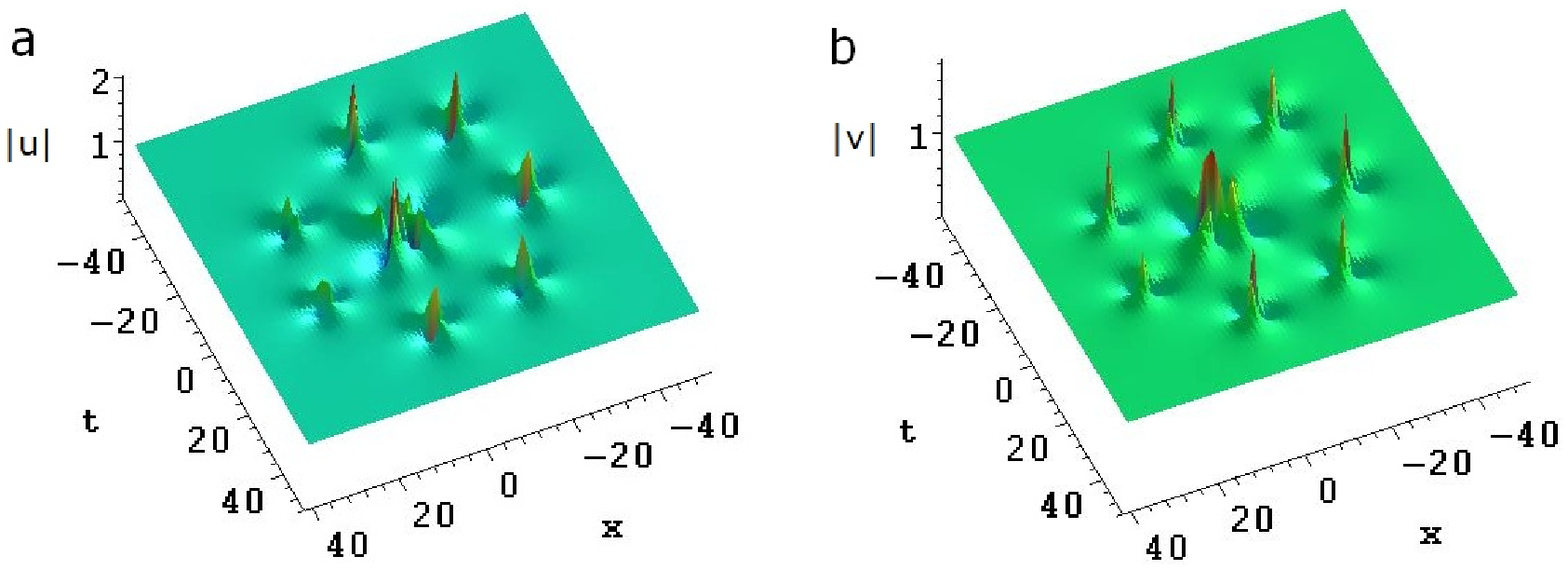}}
\caption{Evolution plots of the third-order rogue waves of circular pattern \Rmnum{2}
in CH equations by choosing $\epsilon=0.01, f_{1}=0,g_{1}=0,h_{1}=0.01,
f_{2}=0,g_{2}=0,h_{2}=0,f_{3}=0,g_{3}=1000000,h_{3}=0$.
(a) In $u$ component; (b) in $v$ component.}
\end{figure}

\begin{figure}[!h]
\centering
\renewcommand{\figurename}{{\bf Fig.}}
{\includegraphics[height=4.5cm,width=16cm]{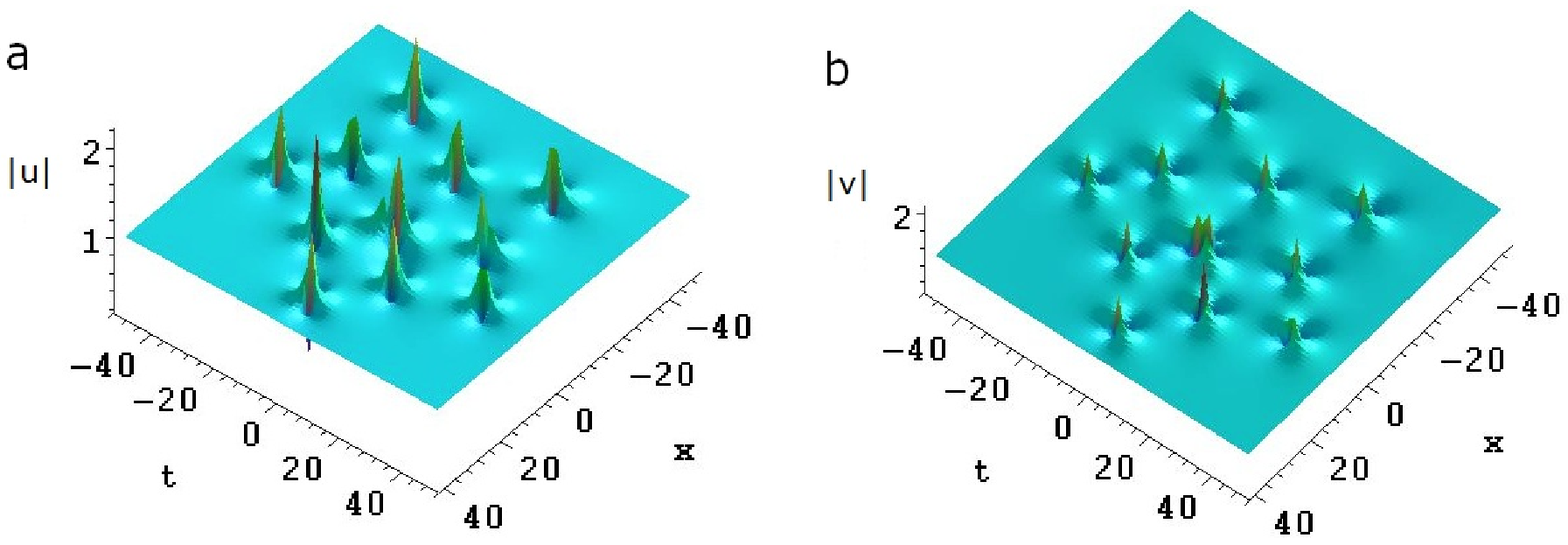}}
\caption{Evolution plots of the third-order rogue waves of pentagram pattern \Rmnum{1}
in CH equations by choosing $\epsilon=0.01, f_{1}=0,g_{1}=0,h_{1}=0.01,
f_{2}=100000,g_{2}=0,h_{2}=0,f_{3}=0,g_{3}=0,h_{3}=0$.
(a) In $u$ component; (b) in $v$ component.}
\centering
\renewcommand{\figurename}{{\bf Fig.}}
{\includegraphics[height=4.5cm,width=16cm]{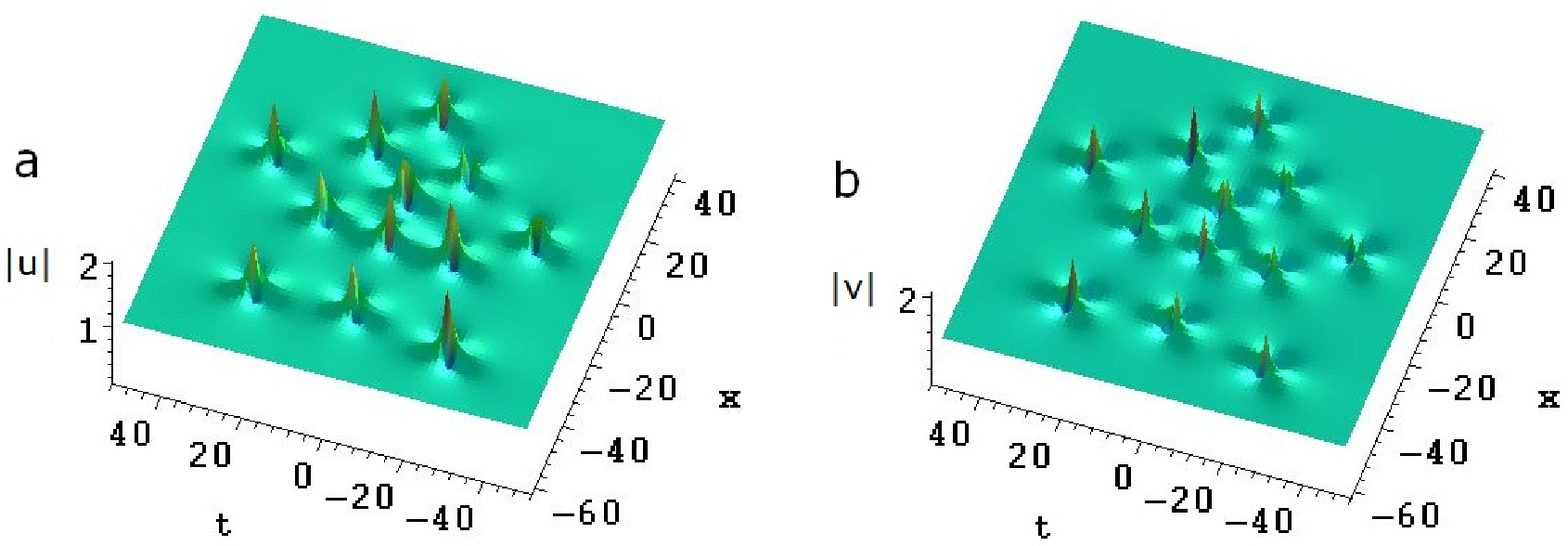}}
\caption{Evolution plots of the third-order rogue waves of pentagram pattern \Rmnum{2}
in CH equations by choosing $\epsilon=0.01, f_{1}=0,g_{1}=1,h_{1}=0.01,
f_{2}=100000,g_{2}=0,h_{2}=0,f_{3}=0,g_{3}=0,h_{3}=0$.
(a) In $u$ component; (b) in $v$ component.}
\centering
\renewcommand{\figurename}{{\bf Fig.}}
{\includegraphics[height=4.5cm,width=16cm]{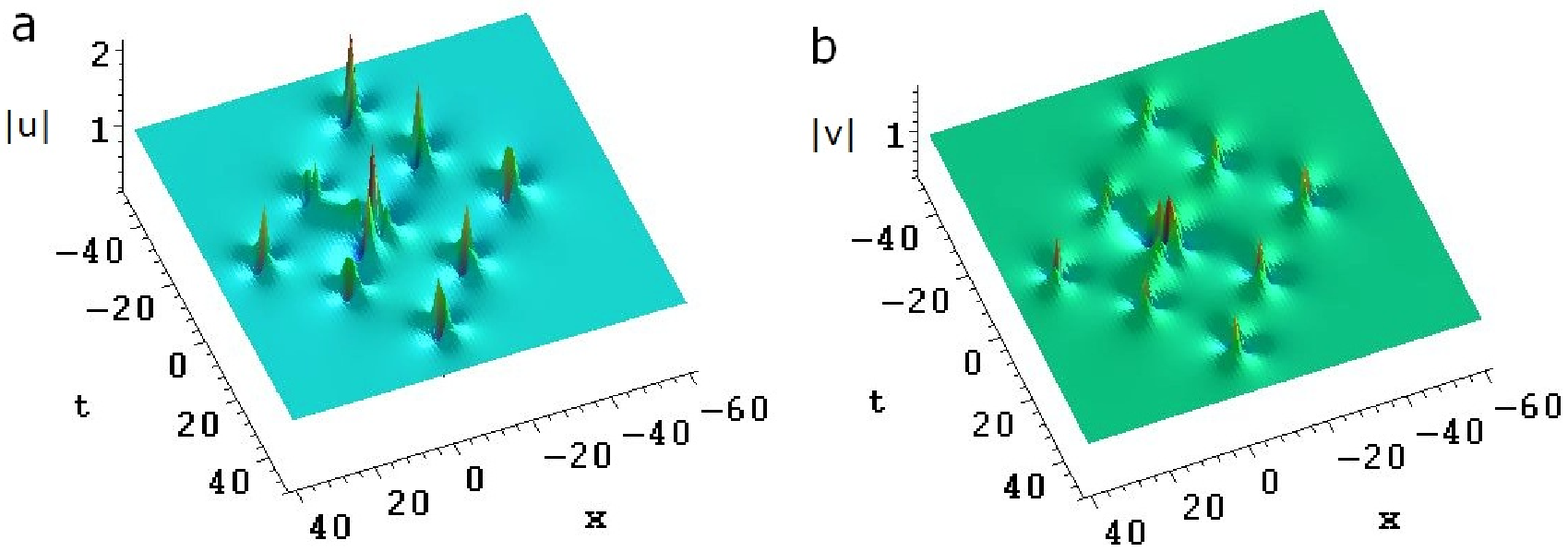}}
\caption{Evolution plots of the third-order rogue waves of quadrilateral pattern \Rmnum{1}
in CH equations by choosing $\epsilon=0.01, f_{1}=0,g_{1}=0,h_{1}=0.01,
f_{2}=0,g_{2}=1000,h_{2}=0,f_{3}=0,g_{3}=0,h_{3}=0$.
(a) In $u$ component; (b) in $v$ component.}
\centering
\renewcommand{\figurename}{{\bf Fig.}}
{\includegraphics[height=4.5cm,width=16cm]{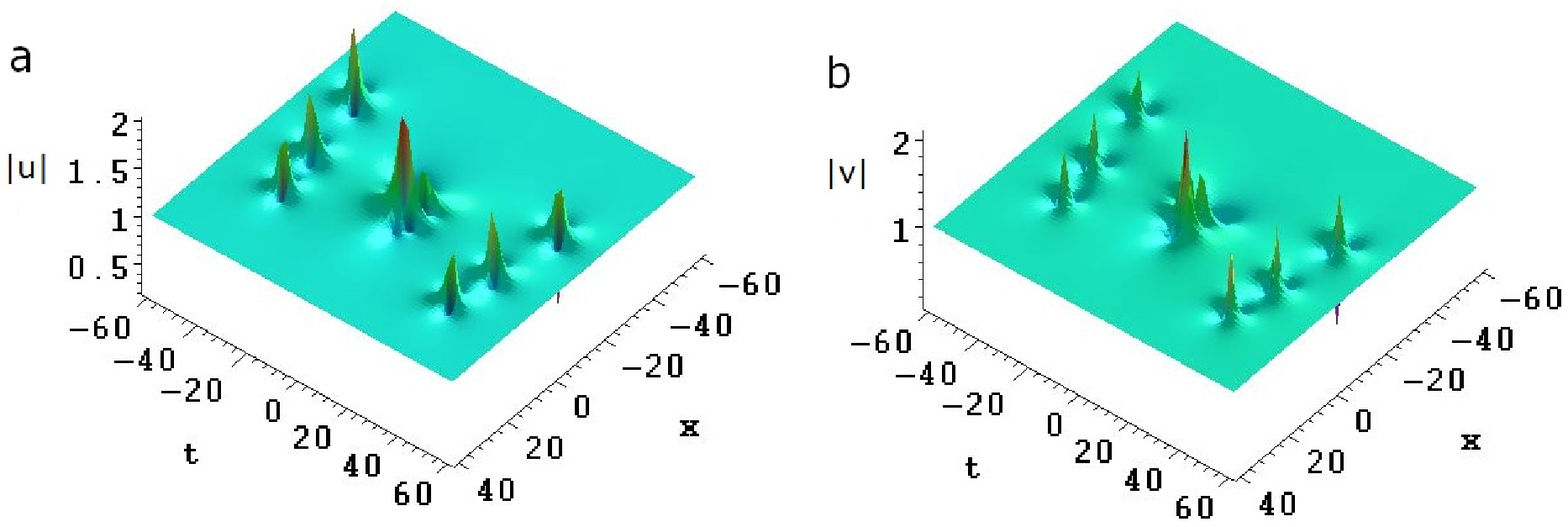}}
\caption{Evolution plots of the third-order rogue waves of quadrilateral pattern \Rmnum{2}
in CH equations by choosing $\epsilon=0.01, f_{1}=100,g_{1}=0,h_{1}=0.01,
f_{2}=0,g_{2}=0,h_{2}=0,f_{3}=0,g_{3}=0,h_{3}=0$.
(a) In $u$ component; (b) in $v$ component.}
\end{figure}

\begin{figure}[!h]
\centering
\renewcommand{\figurename}{{\bf Fig.}}
{\includegraphics[height=4.5cm,width=16cm]{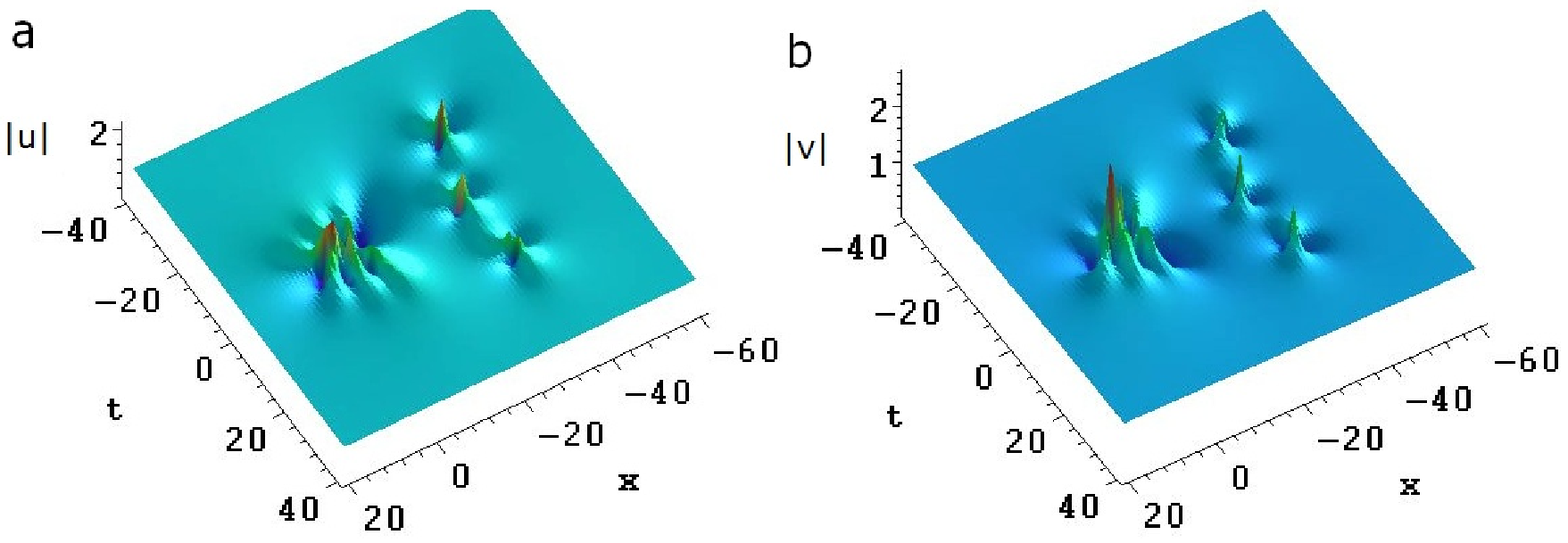}}
\caption{Evolution plots of the third-order rogue waves of triangular pattern
in CH equations by choosing $\epsilon=0.01, f_{1}=0,g_{1}=1,h_{1}=0.01,
f_{2}=0,g_{2}=0,h_{2}=0,f_{3}=0,g_{3}=0,h_{3}=0$.
(a) In $u$ component; (b) in $v$ component.}
\centering
\renewcommand{\figurename}{{\bf Fig.}}
{\includegraphics[height=4.5cm,width=16cm]{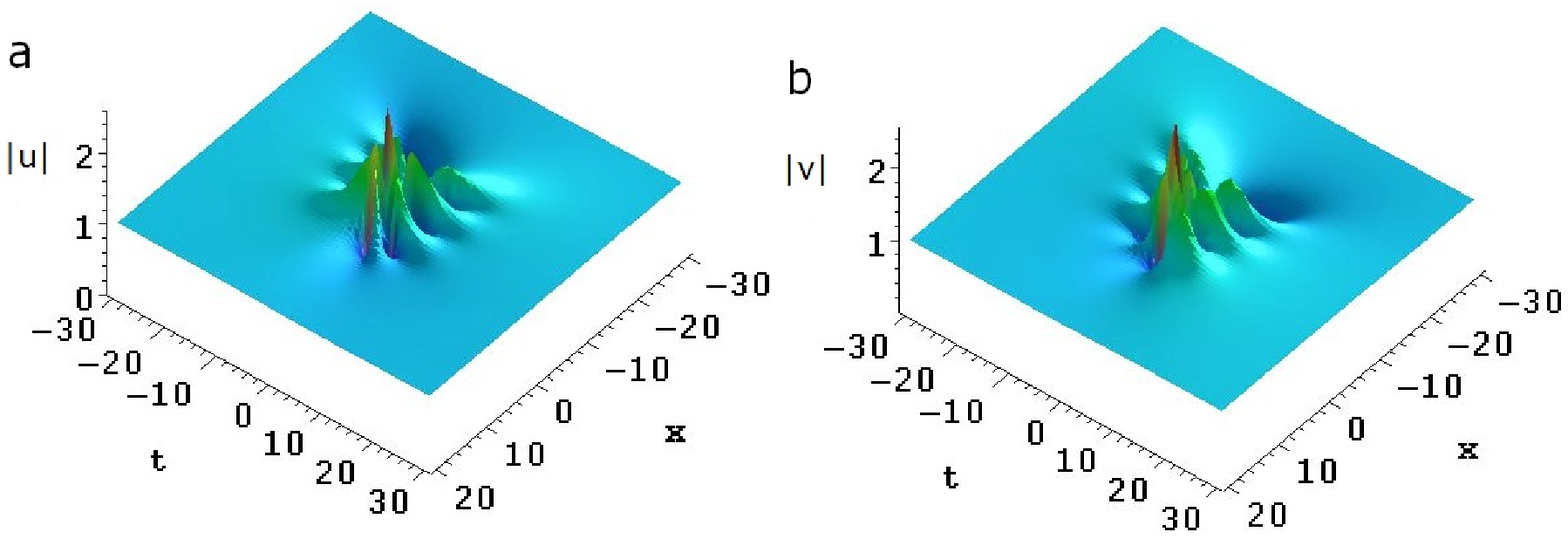}}
\caption{Evolution plots of the third-order rogue waves of fundamental pattern
in CH equations by choosing $\epsilon=0.01, f_{1}=0,g_{1}=0,h_{1}=0.01,
f_{2}=0,g_{2}=0,h_{2}=0,f_{3}=0,g_{3}=0,h_{3}=0$.
(a) In $u$ component; (b) in $v$ component.}
\end{figure}

\begin{table}[!h]
\begin{center}
\caption{Maximum amplitude of the peaks in the third-order rogue waves: twelve fundamental rogue waves case}
\begin{tabular}{lcll}
\toprule
pattern  &  composite no.   & ~~~~~~~~~~~~~~~~~~~~~$|u|$ & ~~~~~~~~~~~~~~~~~~~~~$|v|$  \\
\midrule
pentagram \Rmnum{1} & 2  &  $2.2582,(x=-1.0564,t=-0.1115)$ & $2.2491,(x=-1.0318,t=0.0980)$ \\
quadrilateral \Rmnum{1} & 4  &  $3.0404,(x=-0.8105,t=0.0078)$ & $3.0460,(x=-0.8192,t=-0.0202)$ \\
circular \Rmnum{1} & 4 &  $3.2697,(x=-0.7399,t=-0.0783)$  & $3.2530,(x=-0.7221,t=0.0753)$\\
circular \Rmnum{2} & 5  &  $3.5777,(x=-0.6806,t=0.0538)$ & $3.6008,(x=-0.6958,t=-0.0572)$ \\
quadrilateral \Rmnum{2} & 6  &  $3.7724,(x=-0.6096,t=-0.0014)$ & $3.7751,(x=-0.6102,t=-0.0062)$ \\
triangular & 9  &  $4.1914,(x=-0.5956,t=-0.0380)$ & $4.1442,(x=-0.5845,t=0.0328)$ \\
fundamental & 12  &  $4.8530,(x=-0.5130,t=-0.0023)$ & $4.8558,(x=-0.5134,t=-0.0040)$ \\
\bottomrule
\end{tabular}\\
\end{center}
\end{table}

\section{Conclusion}
In summary, we derive a unified representation of $N$th-order rogue wave solution with $3N+1$ free parameters of the CH equations
via the gDT method. Apart from the first-order composite rogue wave or rogue-wave pair solution,
we devote to exploring higher-order rogue wave solutions of the CH equations.
The explicit second-order rogue wave solution containing polynomial of eighth order is presented,
which is impossible to be obtained for the scalar system. We show that four fundamental rogue waves with
quadrilateral, triangular, line and fundamental patterns can emerge on the
spatial-temporal distribution by choosing different values of the free parameters.
Also, we exhibit that six fundamental rogue waves with the more interesting patterns can
coexist in the second-order rogue waves, and the corresponding solutions are made up of
polynomials of twelfth order.  Moreover, the third-order rogue wave solution consists of
polynomials of 18th or 24th order. So, nine or twelve fundamental rogue waves with the more
intricate composite structures can synchronously emerge in the higher-order rogue waves.
All the solutions computed in this paper have been verified by putting them back into the CH equations,
and when taking $\epsilon\rightarrow0$ they can be reduced to the solutions of the Manakov system.
Further, several interesting wave characteristics such as the maximum amplitudes and the coordinate positions of the
peaks in the rogue waves are given by the numerical computation, and the perturbation influences
produced by the high-order nonlinear effects in the CH equations are also discussed.
Our results may help to better enucleate the dynamics of the complex
rogue wave phenomena governed by the CH equations in the deep ocean and nonlinear optics,
and we hope they will be verified in real experiments in the near future.

In addition, on the one hand, the complete classification of the rogue wave solutions for the coupled
equations have not yet been solved, the fourth, the fifth and even the $N$th-order rogue waves
can present the more complicated composite structures; on the other hand, higher-order
rogue wave solutions of the other coupled systems such as resonance interaction system and
multi-component NLS-type system can also be obtained through the gDT method. Both of
these problems deserve to be further investigated, and we will give
the corresponding results in the future papers.

\section*{Acknowledgment}
The project is supported by the Global Change Research Program of China (No.2015CB953904),
National Natural Science Foundation of China (Grant No.
11275072 and 11435005), Research Fund for the Doctoral Program of Higher Education of China (No. 20120076110024),
Innovative Research Team Program of the National Natural Science Foundation of China (Grant No. 61321064),
Shanghai Knowledge Service Platform for Trustworthy Internet of Things under Grant No. ZF1213,
Shanghai Minhang District talents of high level scientific research project,
Talent Fund and K.C. Wong Magna Fund in Ningbo University.

\section*{Appendix A. Explicit expressions of coefficients in (\ref{28})}
$$\begin{array}{l}
\psi_{1}^{[0]}=[-96h_{1}x^{2}-8\sqrt{3}(g_{1}+32h_{1})x-18h_{1}(363\epsilon^{2}-4)t^{2}+
(1584h_{1}\epsilon x+6\sqrt{3}(11g_{1}+400h_{1})\epsilon)t-2f_{1}-24g_{1}\\~~~~~~-384h_{1}
+{\rm i}(792\sqrt{3}\epsilon h_{1}t^{2}-(96\sqrt{3}h_{1}x+12g_{1}+480h_{1})t )]\exp[{\rm i}\displaystyle\frac{5}{4}t],\\
\phi_{1}^{[0]}=({\rm i}+\sqrt{3})[-48h_{1}x^{2}-4\sqrt{3}(g_{1}+20h_{1})x-9h_{1}(363\epsilon^{2}-4)t^{2}+
(792\epsilon h_{1} x+3\sqrt{3}(11\epsilon g_{1}+268\epsilon h_{1}-8h_{1}))t\\~~~~~~-f_{1}-6g_{1}-48h_{1}+{\rm i}(48h_{1}x+396\sqrt{3}\epsilon h_{1}t^{2}-(48\sqrt{3}h_{1}x+396\epsilon h_{1}+6g_{1}+168h_{1})t+2\sqrt{3}(g_{1}+8h_{1})
)]\\~~~~~~\times\exp[{\rm i}(-\displaystyle\frac{1}{2}x-\frac{(5-23\epsilon)}{8}t)],\\
\chi_{1}^{[0]}=({\rm i}-\sqrt{3})[48h_{1}x^{2}+4\sqrt{3}(g_{1}+20h_{1})x+9h_{1}(363\epsilon^{2}-4)t^{2}
-(792\epsilon h_{1}x+3\sqrt{3}(11\epsilon g_{1}+268\epsilon h_{1}+8h_{1}))t\\~~~~~~+f_{1}+6g_{1}+48h_{1}
+{\rm i}(48h_{1}x-396\sqrt{3}\epsilon h_{1}t^{2}+(48\sqrt{3}h_{1}x-396\epsilon h_{1}+6g_{1}+168h_{1})t+2\sqrt{3}(g_{1}+8h_{1})  )]\\~~~~~~\times\exp[{\rm i}(\displaystyle\frac{1}{2}x-\frac{(5+23\epsilon)}{8}t)],\\

\psi_{1}^{[1]}=-\displaystyle\frac{\sqrt{3}}{7680}[18432h_{1}x^{5}+3840\sqrt{3}(g_{1}+56h_{1})x^{4}+(3840f_{1}
+92160g_{1}+2211840h_{1})x^{3}+1280\sqrt{3}(15f_{1}+156g_{1}\\~~~~~~+2432h_{1}
+192h_{2})x^{2}+(69120f_{1}+481280g_{1}+5488640h_{1}+61440g_{2}+1966080h_{2})x-5346\epsilon h_{1}(131769\epsilon^{4}\\~~~~~~-14520\epsilon^{2}+80)t^{5}+(3240h_{1}(131769\epsilon^{4}-8712\epsilon^{2}+16)x
+135\sqrt{3}(131769\epsilon^{4}g_{1}+9679032\epsilon^{4}h_{1}-8712\epsilon^{2}g_{1}\\~~~~~~-703296\epsilon^{2}h_{1}
+16g_{1}+1408h_{1}))t^{4}+(-855360\epsilon h_{1}(11\epsilon-2)(11\epsilon+2)x^{2}-6480\sqrt{3}(1331\epsilon^{2}g_{1}+91960\epsilon^{2}h_{1}
\\~~~~~~-44g_{1}-3360h_{1})\epsilon x-1620\epsilon(1331\epsilon^{2}f_{1}+40656g_{1}\epsilon^{2}+1153152\epsilon^{2}h_{1}-44f_{1}-1504g_{1}-46208h_{1}))t^{3}
\\~~~~~~+(34560h_{1}(363\epsilon^{2}-4)x^{3}+4320\sqrt{3}(363\epsilon^{2}g_{1}
+23496\epsilon^{2}h_{1}-4g_{1}-288h_{1})x^{2}+(784080\epsilon^{2}f_{1}+22239360\epsilon^{2}g_{1}\\~~~~~~+595952640\epsilon^{2}h_{1}
-8640f_{1}-276480g_{1}-8017920h_{1})x+240\sqrt{3}(6633\epsilon^{2}f_{1}+76932\epsilon^{2}g_{1}+
+1446144\epsilon^{2}h_{1}\\~~~~~~+69696\epsilon^{2}h_{2}-84f_{1}-1056g_{1}-20096h_{1}-768h_{2}))t^{2}
+(-760320\epsilon h_{1} x^{4}-11520\sqrt{3}(11g_{1}+664h_{1})\epsilon x^{3}\\~~~~~~-8640\epsilon(11f_{1}+288g_{1}+7296h_{1})x^{2}-1920\sqrt{3}(183f_{1}+2004g_{1}
+34816h_{1}+2112h_{2})\epsilon x-1920\epsilon(351f_{1}\\~~~~~~+2860g_{1}+41632h_{1}+264g_{2}+9600h_{2}))t+
2560\sqrt{3}(9f_{1}+44g_{1}+352h_{1}+2f_{2}+24g_{2}+384h_{2})
\\~~~~~~+{\rm i}(324\sqrt{3}h_{1}(658845\epsilon^{4}-14520\epsilon^{2}+16)t^{5}+(95040\sqrt{3}h_{1}\epsilon(33\epsilon+2\sqrt{3})(-33\epsilon
+2\sqrt{3})x-3240\epsilon(3993g_{1}\epsilon^{2}\\~~~~~~+307824h_{1}\epsilon^{2}-3712h_{1}-44g_{1}))t^{4}
+(17280\sqrt{3}h_{1}(33\epsilon-2)(33\epsilon+2)x^{2}+(4704480\epsilon^{2}g_{1}+342144000\epsilon^{2}h_{1}\\~~~~~~-17280g_{1}
-1382400h_{1})x+360\sqrt{3}(1089\epsilon^{2}f_{1}+35244\epsilon^{2}g_{1}+1039968\epsilon^{2}h_{1}-4f_{1}-144g_{1}-4608h_{1}))t^{3}
\\~~~~~~+(-1520640\sqrt{3}\epsilon h_{1} x^{3}-51840\epsilon(11g_{1}+752h_{1})x^{2}
-8640\sqrt{3}\epsilon(11f_{1}+332g_{1}+9248h_{1}) x-11520\epsilon(54f_{1}\\~~~~~~+651g_{1}+12304h_{1}+528h_{2}))t^{2}+
(46080\sqrt{3}h_{1}x^{4}+(23040g_{1}+1474560h_{1})x^{3}+5760\sqrt{3}(f_{1}+28g_{1}\\~~~~~~+736h_{1})x^{2}
+(69120f_{1}+783360g_{1}+13762560h_{1}+737280h_{2})x+5120\sqrt{3}(9f_{1}+74g_{1}+1048h_{1}+6g_{2}\\~~~~~~+240h_{2}))t
)]\exp[{\rm i}\displaystyle\frac{5}{4}t],\\

\phi_{1}^{[1]}=-\displaystyle\frac{(\sqrt{3}{\rm i}+3)}{15360}[18432h_{1}x^{5}+3840\sqrt{3}(g_{1}+44h_{1})x^{4}+(3840f_{1}+69120g_{1}+1105920h_{1})x^{3}
+640\sqrt{3}(21f_{1}\\~~~~~~+132g_{1}+1408h_{1}+384h_{2})x^{2}+(23040f_{1}+112640g_{1}+942080h_{1}+61440g_{2}+1228800h_{2})x
\\~~~~~~-5346\epsilon h_{1}(131769\epsilon^{4}-14520\epsilon^{2}+80)t^{5}+
(3240h_{1}(131769\epsilon^{4}-8712\epsilon^{2}+16)x+135\sqrt{3}(
131769\epsilon^{4}g_{1}\\~~~~~~+8097804\epsilon^{4}h_{1}-383328\epsilon^{3}h_{1}-8712\epsilon^{2}g_{1}-598752\epsilon^{2}h_{1}
+4224\epsilon h_{1}+16g_{1}+1216h_{1}))t^{4}+
(-855360\epsilon h_{1}(11\epsilon
\end{array}$$

$$\begin{array}{l}
\\~~~~~~-2)(11\epsilon+2)x^{2}-2160\sqrt{3}(3993\epsilon^{3}g_{1}+227964\epsilon^{3}h_{1}
-8712\epsilon^{2}h_{1}-132\epsilon g_{1}-8496\epsilon h_{1}+32h_{1})x
-2156220\\~~~~~~\times\epsilon^{3}f_{1}-52925400\epsilon^{3}g_{1}-1077753600\epsilon^{3}h_{1}
+2352240\epsilon^{2}g_{1}+114618240\epsilon^{2}h_{1}+71280f_{1}\epsilon+2008800\epsilon g_{1}\\~~~~~~+45619200\epsilon h_{1}-8640g_{1}
-483840h_{1})t^{3}+

(34560h_{1}(363\epsilon^{2}-4)x^{3}+4320\sqrt{3}(363\epsilon^{2}g_{1}+
19140\epsilon^{2}h_{1}-528\epsilon h_{1}\\~~~~~~-4g_{1}-240h_{1})x^{2}
+(784080\epsilon^{2}f_{1}+17534880\epsilon^{2}g_{1}+329080320\epsilon^{2}h_{1}-570240\epsilon g_{1}
-25297920\epsilon h_{1}-8640f_{1}\\~~~~~~-224640g_{1}-4700160h_{1})x
+120\sqrt{3}(9999\epsilon^{2}f_{1}+74268\epsilon^{2}g_{1}+1150464\epsilon^{2}h_{1}
+139392\epsilon^{2}h_{2}-396\epsilon f_{1}\\~~~~~~-7200\epsilon g_{1}-96768\epsilon h_{1}
-132f_{1}-1104g_{1}-16000h_{1}-1536h_{2}))t^{2}+

(-760320\epsilon h_{1} x^{4}
-11520\sqrt{3}(11\epsilon g_{1}\\~~~~~~+532\epsilon h_{1}-8h_{1})x^{3}
+(-95040\epsilon f_{1}-1918080\epsilon g_{1}-33177600\epsilon h_{1}+
34560g_{1}+1382400h_{1})x^{2}
-960\sqrt{3}(267\epsilon f_{1}\\~~~~~~+1812\epsilon g_{1}+24704\epsilon h_{1}+4224\epsilon h_{2}
-6f_{1}-96g_{1}-1152h_{1})x-241920\epsilon f_{1}-1827840\epsilon g_{1}-25835520\epsilon h_{1}\\~~~~~~-506880\epsilon g_{2}
-12349440\epsilon h_{2}+17280f_{1}+69120g_{1}+1167360h_{1}+368640h_{2})t+
1280\sqrt{3}(3f_{1}+8g_{1}+32h_{1}\\~~~~~~+4f_{2}+24g_{2}+192h_{2})

+{\rm i}(-46080h_{1}x^{4}-7680\sqrt{3}(g_{1}+32h_{1})x^{3}-(5760f_{1}+69120g_{1}+737280h_{1})x^{2}
\\~~~~~~-2560\sqrt{3}(3f_{1}+12g_{1}+112h_{1}+96h_{2})x+
324\sqrt{3}h_{1}(658845\epsilon^{4}-14520\epsilon^{2}+16)t^{5}+
(95040\sqrt{3}\epsilon h_{1}(33\epsilon\\~~~~~~+2\sqrt{3})(-33\epsilon+2\sqrt{3})x
-213465780\epsilon^{4}h_{1}-12937320\epsilon^{3}g_{1}-842101920\epsilon^{3}h_{1}
+142560\epsilon g_{1}+14113440\epsilon^{2}h_{1}
\\~~~~~~+10316160\epsilon h_{1}-25920h_{1})t^{4}+

(17280\sqrt{3}h_{1}(33\epsilon-2)(33\epsilon+2)x^{2}
+(103498560\epsilon^{3}h_{1}+4704480\epsilon^{2}g_{1}\\~~~~~~+285690240\epsilon^{2}h_{1}
-3421440\epsilon h_{1}-17280g_{1}-1175040h_{1})x
+360\sqrt{3}(11979\epsilon^{3}g_{1}+540144\epsilon^{3}h_{1}+1089\epsilon^{2}f_{1}\\~~~~~~+28710\epsilon^{2}g_{1}
+617040\epsilon^{2}h_{1}-396\epsilon g_{1}-20736\epsilon h_{1}-4f_{1}-120g_{1}-2880h_{1}))t^{3}+

(-1520640\sqrt{3}\epsilon h_{1} x^{3}
\\~~~~~~+(-18817920\epsilon^{2}h_{1}-570240\epsilon g_{1}-32140800\epsilon h_{1}+
207360h_{1})x^{2}
-4320\sqrt{3}(363\epsilon^{2}g_{1}+14784\epsilon^{2}h_{1}+22\epsilon f_{1}

\\~~~~~~+532\epsilon g_{1}+10528\epsilon h_{1}-4g_{1}-192h_{1})x
-392040\epsilon^{2}f_{1}-6415200\epsilon^{2}g_{1}-81285120\epsilon^{2}h_{1}
-479520\epsilon f_{1}\\~~~~~~-3767040\epsilon g_{1}-56309760\epsilon h_{1}
-6082560\epsilon h_{2}+4320f_{1}+86400g_{1}+1244160h_{1})t^{2}+
(46080\sqrt{3}h_{1}x^{4}\\~~~~~~+(1520640\epsilon h_{1}+23040g_{1}+1198080h_{1})x^{3}
+5760\sqrt{3}(33\epsilon g_{1}+1200\epsilon h_{1}+f_{1}+22g_{1}+400h_{1})x^{2}
\\~~~~~~+(95040\epsilon f_{1}+1347840\epsilon g_{1}+15482880\epsilon h_{1}
+51840f_{1}+368640g_{1}+4915200h_{1}+737280h_{2})x
\\~~~~~~+640\sqrt{3}(126\epsilon f_{1}+504\epsilon g_{1}+8880\epsilon h_{1}+3168\epsilon h_{2}
+27f_{1}+196g_{1}+2528h_{1}+48g_{2}+1344h_{2}))t
-3840f_{1}\\~~~~~~-10240g_{1}-40960h_{1}-30720g_{2}-245760h_{2}
) ]\exp[{\rm i}(-\displaystyle\frac{1}{2}x-\frac{(5-23\epsilon)}{8}t)],\\

\chi_{1}^{[1]}=\displaystyle\frac{(\sqrt{3}{\rm i}-3)}{15360}[18432h_{1}x^{5}+3840\sqrt{3}(g_{1}+44h_{1})x^{4}+(3840f_{1}+69120g_{1}+1105920h_{1})x^{3}
+640\sqrt{3}(21f_{1}\\~~~~~~+132g_{1}+1408h_{1}+384h_{2})x^{2}+(23040f_{1}+112640g_{1}+942080h_{1}+61440g_{2}+1228800h_{2})x
\\~~~~~~-5346h_{1}\epsilon(131769\epsilon^{4}-14520\epsilon^{2}+80)t^{5}+

(3240h_{1}(131769\epsilon^{4}-8712\epsilon^{2}+16)x
+135\sqrt{3}(131769\epsilon^{4}g_{1}\\~~~~~~+809780\epsilon^{4}4h_{1}+383328\epsilon^{3}h_{1}
-8712\epsilon^{2}g_{1}-598752\epsilon^{2}h_{1}-4224\epsilon h_{1}+16g_{1}+1216h_{1}))t^{4}+

(-855360\epsilon h_{1}(11\epsilon\\~~~~~~-2)(11\epsilon+2)x^{2}-2160\sqrt{3}(+3993\epsilon^{3}g_{1}
+227964\epsilon^{3}h_{1}+8712\epsilon^{2}h_{1}-132\epsilon g_{1}-8496\epsilon h_{1}
-32h_{1})x-2156220\\~~~~~~\times\epsilon^{3}f_{1}-52925400\epsilon^{3}g_{1}-1077753600\epsilon^{3}h_{1}
-2352240\epsilon^{2}g_{1}-114618240\epsilon^{2}h_{1}
+71280\epsilon f_{1}+2008800\epsilon g_{1}
\\~~~~~~+45619200\epsilon h_{1}+8640g_{1}+483840h_{1})t^{3}+

(34560h_{1}(363\epsilon^{2}-4)x^{3}
+4320\sqrt{3}(363\epsilon^{2}g_{1}+19140\epsilon^{2}h_{1}
+528\epsilon h_{1}\\~~~~~~-4g_{1}-240h_{1})x^{2}
+(784080\epsilon^{2}f_{1}+17534880\epsilon^{2}g_{1}+329080320\epsilon^{2}h_{1}
+570240\epsilon g_{1}+25297920\epsilon h_{1}-8640f_{1}\\~~~~~~-224640g_{1}-4700160h_{1})x
+120\sqrt{3}(9999\epsilon^{2}f_{1}+74268\epsilon^{2}g_{1}+1150464\epsilon^{2}h_{1}
+139392\epsilon^{2}h_{2}+396\epsilon f_{1}\\~~~~~~+7200\epsilon g_{1}+96768\epsilon h_{1}
-132f_{1}-1104g_{1}-16000h_{1}
-1536h_{2}))t^{2}+

(-760320\epsilon h_{1}x^{4}
-11520\sqrt{3}(11\epsilon g_{1}\\~~~~~~+532\epsilon h_{1}+8h_{1})x^{3}
+(-95040\epsilon f_{1}-1918080\epsilon g_{1}-33177600\epsilon h_{1}
-34560g_{1}-1382400h_{1})x^{2}
-960\sqrt{3}(267\epsilon f_{1}\\~~~~~~+1812\epsilon g_{1}+24704\epsilon h_{1}+4224\epsilon h_{2}
+6f_{1}+96g_{1}+1152h_{1})x
-241920\epsilon f_{1}-1827840\epsilon g_{1}-25835520\epsilon h_{1}
\\~~~~~~-506880\epsilon g_{2}-12349440\epsilon h_{2}-17280f_{1}-69120g_{1}-1167360h_{1}-368640h_{2})t+
1280\sqrt{3}(3f_{1}+8g_{1}+32h_{1}\\~~~~~~+4f_{2}+24g_{2}+192h_{2})

+{\rm i}(46080h_{1}x^{4}+7680\sqrt{3}(g_{1}+32h_{1})x^{3}+(5760f_{1}+69120g_{1}+737280h_{1})x^{2}\\~~~~~~+2560\sqrt{3}(3f_{1}+12g_{1}
+112h_{1}+96h_{2})x+324\sqrt{3}h_{1}(658845\epsilon^{4}-14520\epsilon^{2}+16)t^{5}+

(95040\sqrt{3}\epsilon h_{1}(33\epsilon\\~~~~~~+2\sqrt{3})(-33\epsilon+2\sqrt{3})x+
213465780\epsilon^{4}h_{1}-12937320\epsilon^{3}g_{1}-842101920\epsilon^{3}h_{1}
-14113440\epsilon^{2}h_{1}+142560\epsilon g_{1}\\~~~~~~+10316160\epsilon h_{1}+25920h_{1})t^{4}

(17280\sqrt{3}h_{1}(33\epsilon-2)(33\epsilon+2)x^{2}
+(-103498560\epsilon^{3}h_{1}+4704480\epsilon^{2}g_{1}\\~~~~~~+285690240\epsilon^{2}h_{1}
+3421440\epsilon h_{1}-17280g_{1}-1175040h_{1})x
+360\sqrt{3}(-11979\epsilon^{3}g_{1}-540144\epsilon^{3}h_{1}+1089\epsilon^{2}f_{1}\\~~~~~~+28710\epsilon^{2}g_{1}
+617040\epsilon^{2}h_{1}+396\epsilon g_{1}+20736\epsilon h_{1}-4f_{1}-120g_{1}
-2880h_{1}))t^{3}+

(-1520640\sqrt{3}\epsilon h_{1} x^{3}
\\~~~~~~+(18817920\epsilon^{2}h_{1}-570240\epsilon g_{1}-32140800\epsilon h_{1}
-207360h_{1})x^{2}
-4320\sqrt{3}(-363\epsilon^{2}g_{1}-14784\epsilon^{2}h_{1}+22\epsilon f_{1}\\~~~~~~+532\epsilon g_{1}
+10528\epsilon h_{1}+4g_{1}+192h_{1})x
+392040\epsilon^{2}f_{1}+6415200\epsilon^{2}g_{1}+81285120\epsilon^{2}h_{1}
-479520\epsilon f_{1}\\~~~~~~-3767040\epsilon g_{1}-56309760\epsilon h_{1}-6082560\epsilon h_{2}
-4320f_{1}-86400g_{1}-1244160h_{1})t^{2}+
(46080\sqrt{3}h_{1}x^{4}

\\~~~~~~+(1520640h_{1}\epsilon+23040g_{1}+1198080h_{1})x^{3}
+5760\sqrt{3}(-33\epsilon g_{1}-1200\epsilon h_{1}+f_{1}+22g_{1}+400h_{1})x^{2}
\\~~~~~~+(-95040\epsilon f_{1}-1347840\epsilon g_{1}-15482880\epsilon h_{1}+51840f_{1}
+368640g_{1}+4915200h_{1}+737280h_{2})x\\~~~~~~-640\sqrt{3}(126\epsilon f_{1}+504\epsilon g_{1}
+8880\epsilon h_{1}+3168\epsilon h_{2}-27f_{1}-196g_{1}-2528h_{1}-48g_{2}-1344h_{2}))t+
3840f_{1}\\~~~~~~+10240g_{1}+40960h_{1}+30720g_{2}+245760h_{2}
)]\exp[{\rm i}(\displaystyle\frac{1}{2}x-\frac{(5+23\epsilon)}{8}t)].
\end{array}
$$

\section*{Appendix B. Explicit expressions of terms in (\ref{30})}
$$\begin{array}{l}
F_{2}=5308416x^{8}+42467328\sqrt{3}x^{7}+371589120x^{6}+497811456\sqrt{3}x^{5}
-46327136256x^{4}-188746825728\sqrt{3}x^{3}\\~~~~~-771423731712x^{2}-535040098304\sqrt{3}x

+6561(363\epsilon^{2}+4)^{4}t^{8}+(-2309472\epsilon(363\epsilon^{2}+4)^{3}x
-69984\sqrt{3}\\~~~~~\times(14157\epsilon^{2}+196)\epsilon(363\epsilon^{2}+4)^{2})t^{7}+

(139968(2541\epsilon^{2}+4)(363\epsilon^{2}+4)^{2}x^{2}+93312\sqrt{3}(363\epsilon^{2}+4)(3198393\epsilon^{4}
\\~~~~~+47784\epsilon^{2}+80)x+152573385460320\epsilon^{6}-3651801791616\epsilon^{5}+4169234374272\epsilon^{4}
-80480480256\epsilon^{3}\\~~~~~+34513869312\epsilon^{2}-443418624\epsilon+88086528)t^{6}+

(-36951552\epsilon(363\epsilon^{2}+4)(847\epsilon^{2}+4)x^{3}-1119744\sqrt{3}\epsilon\\~~~~~\times(34391709\epsilon^{4}
+603064\epsilon^{2}+2256)x^{2}+
(-106509208912128\epsilon^{5}+2213213207040\epsilon^{4}
-1946952640512\epsilon^{3}\\~~~~~+29265629184\epsilon^{2}-7826264064\epsilon+53747712)x
-62208\sqrt{3}(380356119\epsilon^{5}-38093220\epsilon^{4}+5738184\epsilon^{3}
\\~~~~~-608256\epsilon^{2}+9712\epsilon-1728))t^{5}+

((1721388049920\epsilon^{4}+16258682880\epsilon^{2}+17915904)x^{4}
+497664\sqrt{3}\\~~~~~\times(15273225\epsilon^{4}+162888\epsilon^{2}+208)x^{3}
+(30945557592576\epsilon^{4}-536536535040\epsilon^{3}+342596874240\epsilon^{2}\\~~~~~-3547348992\epsilon+443916288)x^{2}
+82944\sqrt{3}(166770351\epsilon^{4}-13695264\epsilon^{3}+1400472\epsilon^{2}-112320\epsilon-208)x

\\~~~~~-215628012195840\epsilon^{4}-1568125329408\epsilon^{3}+14389562400768\epsilon^{2}
-14237171712\epsilon-27245445120)t^{4}

\\~~~~~+
(-197074944\epsilon(847\epsilon^{2}+4)x^{5}-1990656\sqrt{3}\epsilon(451935\epsilon^{2}+2444)x^{4}
+(4789512364032\epsilon^{3}+65034731520\epsilon^{2}\\~~~~~-26937556992\epsilon+143327232)x^{3}
-1990656\sqrt{3}(1618567\epsilon^{3}-102564\epsilon^{2}+6428\epsilon-288)x^{2}
\\~~~~~+(104352695746560\epsilon^{3}+552168161280\epsilon^{2}-3484003663872\epsilon+1656225792)x
+1327104\sqrt{3}(94625495\epsilon^{3}\\~~~~~+113958\epsilon^{2}-3376752\epsilon+232))t^{3}+

((10116513792\epsilon^{2}+15925248)x^{6}
+29196288\sqrt{3}(2187\epsilon^{2}+4)x^{5}
\\~~~~~+(416443244544\epsilon^{2}-3941498880\epsilon+798916608)x^{4}
+2654208\sqrt{3}(140843\epsilon^{2}-6144\epsilon+180)x^{3}
\\~~~~~+(-18946745303040\epsilon^{2}-65261666304\epsilon+210952912896)x^{2}
-1769472\sqrt{3}(24461559\epsilon^{2}+19548\epsilon\\~~~~~-292360)x
-55980320292864\epsilon^{2}-3522622390272\epsilon+623672819712)t^{2}+

(-350355456\epsilon x^{7}\\~~~~~-2516189184\sqrt{3}\epsilon x^{6}+(-19285475328\epsilon+95551488)x^{5}
-1769472\sqrt{3}(12211\epsilon-276)x^{4}\\~~~~~+(1529602375680\epsilon+2590507008)x^{3}
+7077888\sqrt{3}(700579\epsilon+282)x^{2}+(13303980490752\epsilon\\~~~~~+426598465536)x+1572864\sqrt{3}(2417927\epsilon+240084))t+

104417164525568,\\

H_{2}=21233664x^{7}+138018816\sqrt{3}x^{6}+1040449536x^{5}+1325924352\sqrt{3}x^{4}-91606745088x^{3}\\~~~~~-234900946944\sqrt{3}x^{2}
-722982666240x

-104976(11\epsilon+2)(363\epsilon^{2}+4)^{3}t^{7}+

(139968(2541\epsilon^{2}+396\epsilon+4)\\~~~~~\times(363\epsilon^{2}+4)^{2}x+23328\sqrt{3}(363\epsilon^{2}+4)(6001479\epsilon^{4}
+962676\epsilon^{3}+95568\epsilon^{2}+12528\epsilon+208))t^{6}+

(-1679616\\~~~~~\times(363\epsilon^{2}+4)(27951\epsilon^{3}+3630\epsilon^{2}+132\epsilon+8)x^{2}
-559872\sqrt{3}(64435041\epsilon^{5}+8728698\epsilon^{4}+1174184\epsilon^{3}\\~~~~~+129360\epsilon^{2}
+4688\epsilon+288)x-49971671532288\epsilon^{5}-7553784066048\epsilon^{4}-915757996032\epsilon^{3}
\\~~~~~-139385733120\epsilon^{2}-3317428224\epsilon-573308928)t^{5}+

((3442776099840\epsilon^{4}+357691023360\epsilon^{3}\\~~~~~+32517365760\epsilon^{2}+
2364899328\epsilon+35831808)x^{3}
+1119744\sqrt{3}(9523305\epsilon^{4}+1046408\epsilon^{3}+105688\epsilon^{2}\\~~~~~+7968\epsilon+144)x^{2}
+(29009554928640\epsilon^{4}+3474431262720\epsilon^{3}+327520641024\epsilon^{2}+32057524224\epsilon\\~~~~~+378224640)x
+165888\sqrt{3}(45827793\epsilon^{4}+5947236\epsilon^{3}+408204\epsilon^{2}+57696\epsilon-736))t^{4}+

(-(417306193920\epsilon^{3}\\~~~~~+32517365760\epsilon^{2}+1970749440\epsilon+71663616)x^{4}
-1990656\sqrt{3}(843975\epsilon^{3}+70554\epsilon^{2}+4756\epsilon+184)x^{3}
\\~~~~~-(6727941513216\epsilon^{3}+601831047168\epsilon^{2}
+39295549440\epsilon+1847328768)x^{2}
-2654208\sqrt{3}(1322454\epsilon^{3}\\~~~~~+119187\epsilon^{2}+6066\epsilon+376)x
+51301024579584\epsilon^{3}-29073268113408\epsilon^{2}-1751660494848\epsilon
\\~~~~~+105669328896)t^{3}+

((30349541376\epsilon^{2}+1576599552\epsilon+47775744)x^{5}
+663552\sqrt{3}(224235\epsilon^{2}+12684\epsilon\\~~~~~+428)x^{4}
+(779150721024\epsilon^{2}+46565425152\epsilon+1581907968)x^{3}
+5308416\sqrt{3}(114135\epsilon^{2}+6414\epsilon+182)x^{2}
\\~~~~~+(-18664629534720\epsilon^{2}+7033460097024\epsilon+212280016896)x
-5308416\sqrt{3}(3454183\epsilon^{2}-1473168\epsilon\\~~~~~-42148))t^{2}+

(-(1226244096\epsilon+31850496)x^{6}-5308416\sqrt{3}(1323\epsilon+38)x^{5}
-(45052526592\epsilon\\~~~~~+1358954496)x^{4}
-14155776\sqrt{3}(3274\epsilon+87)x^{3}
+(2264464097280\epsilon-425536782336)x^{2}\\~~~~~+14155776\sqrt{3}(293875\epsilon-63292)x+6674519162880\epsilon-1037203144704)t
-251624685568\sqrt{3},\\

D_{2}=-5308416x^{8}-42467328\sqrt{3}x^{7}-424673280x^{6}-773849088\sqrt{3}x^{5}+44543508480x^{4}+186755579904\sqrt{3}x^{3}
\\~~~~~+626213781504x^{2}+250810990592\sqrt{3}x

-6561(363\epsilon^{2}+4)^{4}t^{8}+

(2309472\epsilon(363\epsilon^{2}+4)^{3}x+69984\sqrt{3}\epsilon\\~~~~~\times(14157\epsilon^{2}+196)(363\epsilon^{2}+4)^{2})t^{7}+

(-139968(2541\epsilon^{2}+4)(363\epsilon^{2}+4)^{2}x^{2}-93312\sqrt{3}(363\epsilon^{2}+4)(3198393\epsilon^{4}
\\~~~~~+47784\epsilon^{2}+80)x-169310810338560\epsilon^{6}
-4943858996736\epsilon^{4}-45488480256\epsilon^{2}-137355264)t^{6}\\~~~~~+

(36951552\epsilon(363\epsilon^{2}+4)(847\epsilon^{2}+4)x^{3}+1119744\sqrt{3}\epsilon(
34391709\epsilon^{4}+603064\epsilon^{2}+2256)x^{2}
+5971968\epsilon\\~~~~~\times(19873161\epsilon^{4}+388905\epsilon^{2}+1756)x
+497664\sqrt{3}\epsilon(71411175\epsilon^{4}+1548180\epsilon^{2}+7712))t^{5}\\~~~~~+

((-1721388049920\epsilon^{4}-16258682880\epsilon^{2}-17915904)x^{4}-497664\sqrt{3}(15273225\epsilon^{4}+162888\epsilon^{2}
+208)x^{3}\\~~~~~-(34634246270976\epsilon^{4}+410883342336\epsilon^{2}+605159424)x^{2}
-663552\sqrt{3}(31452201\epsilon^{4}
+403452\epsilon^{2}+640)x\\~~~~~+205401300664320\epsilon^{4}-14660862935040\epsilon^{2}+25899761664)t^{4}+

(197074944\epsilon(847\epsilon^{2}+4)x^{5}
\end{array}$$

$$\begin{array}{l}
\\~~~~~+1990656\sqrt{3}\epsilon(451935\epsilon^{2}+2444)x^{4}
+31850496\epsilon(169092\epsilon^{2}+1019)x^{3}+15925248\sqrt{3}\epsilon(307001\epsilon^{2}+1966)x^{2}
\\~~~~~-785645568\epsilon(126843\epsilon^{2}-4516)x-10616832\sqrt{3}\epsilon(11690269\epsilon^{2}-424176))t^{3}+

((-10116513792\epsilon^{2}\\~~~~~-15925248)x^{6}-29196288\sqrt{3}(2187\epsilon^{2}+4)x^{5}-(470638854144\epsilon^{2}+966131712)x^{4}
-21233664\sqrt{3}(26899\epsilon^{2}

\\~~~~~+58)x^{3}+(18136946442240\epsilon^{2}-214743121920)x^{2}
+14155776\sqrt{3}(3023343\epsilon^{2}-36728)x\\~~~~~+46008869584896\epsilon^{2}-519677411328)t^{2}+

(350355456\epsilon x^{7}+2516189184\sqrt{3}\epsilon x^{6}+21913141248\epsilon x^{5}\\~~~~~+33280229376\sqrt{3}\epsilon x^{4}
-1467557609472\epsilon x^{3}-4904749891584\sqrt{3}\epsilon x^{2}-10898078957568\epsilon x\\~~~~~-1314102706176\sqrt{3}\epsilon)t

-104732102230016,\\

G_{2}=-5308416x^{8}-42467328\sqrt{3}x^{7}-371589120x^{6}-497811456\sqrt{3}x^{5}+46327136256x^{4}
+188746825728\sqrt{3}x^{3}\\~~~~~+771423731712x^{2}+535040098304\sqrt{3}x+

-6561(363\epsilon^{2}+4)^{4}t^{8}+

(2309472\epsilon(363\epsilon^{2}+4)^{3}x+69984\sqrt{3}\\~~~~~\times(14157\epsilon^{2}+196)\epsilon(363\epsilon^{2}+4)^{2})t^{7}+

(-139968(2541\epsilon^{2}+4)(363\epsilon^{2}+4)^{2}x^{2}-93312\sqrt{3}(363\epsilon^{2}+4)(3198393\epsilon^{4}
\\~~~~~+47784\epsilon^{2}+80)x-152573385460320\epsilon^{6}-3651801791616\epsilon^{5}-4169234374272\epsilon^{4}
-80480480256\epsilon^{3}

\\~~~~~-34513869312\epsilon^{2}
-443418624\epsilon-88086528)t^{6}+

(36951552\epsilon(363\epsilon^{2}+4)(847\epsilon^{2}+4)x^{3}+1119744\sqrt{3}\epsilon

\\~~~~~\times(34391709\epsilon^{4}+603064\epsilon^{2}
+2256)x^{2}
+(106509208912128\epsilon^{5}+2213213207040\epsilon^{4}+1946952640512\epsilon^{3}

\\~~~~~+
29265629184\epsilon^{2}+7826264064\epsilon+53747712)x
+62208\sqrt{3}(380356119\epsilon^{5}+38093220\epsilon^{4}+5738184\epsilon^{3}
\\~~~~~+608256\epsilon^{2}+9712\epsilon+1728))t^{5}+

(-(1721388049920\epsilon^{4}+16258682880\epsilon^{2}+17915904)x^{4}
-497664\sqrt{3}\\~~~~~\times(15273225\epsilon^{4}+162888\epsilon^{2}+208)x^{3}
-(30945557592576\epsilon^{4}+536536535040\epsilon^{3}+342596874240\epsilon^{2}\\~~~~~+3547348992\epsilon+443916288
)x^{2}-82944\sqrt{3}(166770351\epsilon^{4}+13695264\epsilon^{3}+1400472\epsilon^{2}+112320\epsilon-208)x
\\~~~~~+215628012195840\epsilon^{4}-1568125329408\epsilon^{3}-14389562400768\epsilon^{2}-14237171712\epsilon+27245445120)t^{4}\\~~~~~+

(197074944\epsilon(847\epsilon^{2}+4)x^{5}+1990656\sqrt{3}\epsilon(451935\epsilon^{2}+2444)x^{4}
+(4789512364032\epsilon^{3}+65034731520\epsilon^{2}\\~~~~~+26937556992\epsilon+143327232)x^{3}
+1990656\sqrt{3}(1618567\epsilon^{3}+102564\epsilon^{2}+6428\epsilon+288)x^{2}
\\~~~~~+(-104352695746560\epsilon^{3}+552168161280\epsilon^{2}+3484003663872\epsilon+1656225792)x
-1327104\sqrt{3}(94625495\epsilon^{3}\\~~~~~-113958\epsilon^{2}-3376752\epsilon-232))t^{3}+

(-(10116513792\epsilon^{2}+15925248)x^{6}-29196288\sqrt{3}(2187\epsilon^{2}+4)x^{5}\\~~~~~-
(416443244544\epsilon^{2}+3941498880\epsilon+798916608)x^{4}
-2654208\sqrt{3}(140843\epsilon^{2}+6144\epsilon+180)x^{3}
\\~~~~~+(18946745303040\epsilon^{2}-65261666304\epsilon-210952912896)x^{2}
+1769472\sqrt{3}(24461559\epsilon^{2}-19548\epsilon\\~~~~~-292360)x
+55980320292864\epsilon^{2}-3522622390272\epsilon-623672819712)t^{2}+

(350355456\epsilon x^{7}\\~~~~~+2516189184\sqrt{3}\epsilon x^{6}+(19285475328\epsilon+95551488)x^{5}
+1769472\sqrt{3}(12211\epsilon+276)x^{4}\\~~~~~+(-1529602375680\epsilon+2590507008)x^{3}-7077888\sqrt{3}(700579\epsilon-282)x^{2}
+(13303980490752\epsilon\\~~~~~+426598465536)x-1572864\sqrt{3}(2417927\epsilon-240084))t

-104417164525568,\\

K_{2}=21233664x^{7}+138018816\sqrt{3}x^{6}+1040449536x^{5}+1325924352\sqrt{3}x^{4}-91606745088x^{3}
\\~~~~~-234900946944\sqrt{3}x^{2}-722982666240x

-104976(11\epsilon-2)(363\epsilon^{2}+4)^{3}t^{7}+

(139968(2541\epsilon^{2}-396\epsilon+4)\\~~~~~\times(363\epsilon^{2}+4)^{2}x+23328\sqrt{3}(363\epsilon^{2}+4)(6001479\epsilon^{4}
-962676\epsilon^{3}+95568\epsilon^{2}-12528\epsilon+208))t^{6}+

(-1679616\\~~~~~\times(363\epsilon^{2}+4)(27951\epsilon^{3}-3630\epsilon^{2}+132\epsilon-8)x^{2}
-559872\sqrt{3}(64435041\epsilon^{5}-8728698\epsilon^{4}+1174184\epsilon^{3}
\\~~~~~-129360\epsilon^{2}+4688\epsilon-288)x
-49971671532288\epsilon^{5}+7553784066048\epsilon^{4}-915757996032\epsilon^{3}
\\~~~~~+139385733120\epsilon^{2}-3317428224\epsilon+573308928)t^{5}+

((3442776099840\epsilon^{4}-357691023360\epsilon^{3}\\~~~~~+32517365760\epsilon^{2}-2364899328\epsilon+35831808)x^{3}
+1119744\sqrt{3}(9523305\epsilon^{4}-1046408\epsilon^{3}+105688\epsilon^{2}\\~~~~~-7968\epsilon+144)x^{2}
+(29009554928640\epsilon^{4}-3474431262720\epsilon^{3}+327520641024\epsilon^{2}-32057524224\epsilon\\~~~~~+378224640)x
+165888\sqrt{3}(45827793\epsilon^{4}-5947236\epsilon^{3}+408204\epsilon^{2}-57696\epsilon-736))t^{4}+

((-417306193920\epsilon^{3}\\~~~~~+32517365760\epsilon^{2}-1970749440\epsilon+71663616)x^{4}
-1990656\sqrt{3}(843975\epsilon^{3}-70554\epsilon^{2}+4756\epsilon-184)x^{3}
\\~~~~~+(-6727941513216\epsilon^{3}+601831047168\epsilon^{2}-39295549440\epsilon+1847328768)x^{2}
-2654208\sqrt{3}(1322454\epsilon^{3}\\~~~~~-119187\epsilon^{2}+6066\epsilon-376)x
+51301024579584\epsilon^{3}+29073268113408\epsilon^{2}-1751660494848\epsilon\\~~~~~-105669328896)t^{3}+

((30349541376\epsilon^{2}-1576599552\epsilon+47775744)x^{5}+663552\sqrt{3}(224235\epsilon^{2}
-12684\epsilon\\~~~~~+428)x^{4}+(779150721024\epsilon^{2}-46565425152\epsilon+1581907968)x^{3}
+5308416\sqrt{3}(114135\epsilon^{2}-6414\epsilon+182)x^{2}\\~~~~~+(-18664629534720\epsilon^{2}-7033460097024\epsilon+212280016896
)x-5308416\sqrt{3}(3454183\epsilon^{2}+1473168\epsilon\\~~~~~-42148))t^{2}+

((-1226244096\epsilon+31850496)x^{6}-5308416\sqrt{3}(1323\epsilon-38)x^{5}+(
-45052526592\epsilon\\~~~~~+1358954496)x^{4}-14155776\sqrt{3}(3274\epsilon-87)x^{3}+(2264464097280\epsilon+425536782336)x^{2}
\\~~~~~+14155776\sqrt{3}(293875\epsilon+63292)x+6674519162880\epsilon+1037203144704)t

-251624685568\sqrt{3}.
\end{array}
$$


\section*{References}
\begin{thebibliography}{99}
\itemsep=-1pt plus.1pt minus.1pt
\small
\bibitem{01}P. M\"{u}ller, C. Garrett, A. Osborne, Oceanography 18 (2005) 66. 
\bibitem{02}C. Kharif, E. Pelinovsky, Eur. J. Mech. B (Fluids) 22 (2003) 603.
\bibitem{04}D.R. Solli, C. Ropers, P. Koonath, B. Jalali,  Nature 450 (2007)  1054.
\bibitem{05}C. Garrett, J. Gemmrich, Phys. Today 62 (2009) 62.
\bibitem{06}N. Akhmediev, J.M. Dudley, D.R. Solli, S.K. Turitsyn, J. Opt. 15 (2013) 060201.
\bibitem{07}A. Chabchoub, M. Fink, Phys. Rev. Lett. 112 (2014) 124101.
\bibitem{08}N. Akhmediev, A. Ankiewicz, M. Taki, Phys. Lett. A 373 (2009) 675. 
\bibitem{09}A. Zaviyalov, O. Egorov, R. Iliew, F. Lederer, Phys.  Rev.  A 85 (2012) 013828. 
\bibitem{10}M. Shats, H. Punzmann, H. Xia,   Phys. Rev. Lett.  104 (2010) 104503. 
\bibitem{11}W.M. Moslem, P.K. Shukla, B. Eliasson,  Euro. Phys. Lett. 96 (2011) 25002. 
\bibitem{12}Y.V. Bludov, V.V. Konotop, N. Akhmediev, Phys. Rev. A  80 (2009)  033610. 
\bibitem{13}A. Chabchoub, N.P. Hoffmann, N. Akhmediev, Phys. Rev. Lett. 106 (2011) 204502. 
\bibitem{14}L. Stenflo, M. Marklund, J. Plasma Phys. 76 (2010) 293. 
\bibitem{15}V.B. Efimov, A.N.Ganshin, G.V. Kolmakov, P.V.E. McClintock, L.P. Mezhov-Deglin, Eur. Phys. J. Spec. Top. 185 (2010) 181.  
\bibitem{16}D.H. Peregrine, J. Austral, Math. Soc. B:  Appl. Math. 25 (1983) 16. 
\bibitem{17}N. Akhmediev, V.M. Eleonskii, N.E. Kulagin, Theor. Math. Phys. 72 (1987) 183. 
\bibitem{18}Y.C. Ma, Stud. Appl. Math. 60 (1979) 43.
\bibitem{19}B. Kibler, J. Fatome, C. Finot,	G. Millot, F. Dias, G. Genty, N. Akhmediev, J.M. Dudley, Nat. Phys. 6 (2010) 790.
\bibitem{20}C. Lecaplain, P. Grelu,  Phys. Rev. A  90 (2014) 013805. 
\bibitem{21}N. Akhmediev, A. Ankiewicz,  J.M. Soto-Crespo, Phys. Rev. E 80 (2009) 026601.
\bibitem{22}D.J. Kedziora, A. Ankiewicz,  N. Akhmediev, Phys. Rev. E 84 (2011) 056611. 
\bibitem{23}N. Akhmediev, J.M. Soto-Crespo, A. Ankiewicz, Phys. Lett. A  373 (2009) 2137. 
\bibitem{24}D.J. Kedziora, A. Ankiewicz,  N. Akhmediev, Phys. Rev. E 88 (2013) 013207. 
\bibitem{25}A. Ankiewicz, D.J. Kedziora, N. Akhmediev, Phys. Lett. A  375 (2011) 2782.  
\bibitem{26}A. Ankiewicz, N. Devine, N. Akhmediev,  Phys. Lett. A 373 (2009) 3997.
\bibitem{28}P. Dubard, P. Gaillard, C. Klein, V.B. Matveev, Eur. Phys. J. Spec. Top. 185 (2010) 247.
\bibitem{31}B.L. Guo, L.M. Ling, Q.P. Liu, Phys. Rev. E 85 (2012) 026607. 
\bibitem{30s}L.M. Ling, L.C. Zhao, Phys. Rev. E  88 (2013) 043201. 
\bibitem{32}J.S. He, H.R. Zhang, L.H. Wang, K. Porsezian, A.S. Fokas, Phys. Rev. E 87  (2013) 052914.
\bibitem{33}B.L. Guo, L.M. Ling, Q.P. Liu, Stud. Appl. Math. 130  (2013) 317. 
\bibitem{34}S.W. Xu, J.S. He,  J. Phys. A  44 (2011) 305203. 
\bibitem{chow}H.N. Chan, K.W. Chow, D.J. Kedziora, R. Grimshaw, E. Ding, Phys. Rev. E 89 (2014) 032914.
\bibitem{36}Y.S. Tao, J.S. He, Phys. Rev. E 85 (2012) 026601. 
\bibitem{37}L.J. Li, Z.W. Wu, L.H. Wang, J.S. He, Ann. Phys. 334 (2013) 198. 
\bibitem{38}U. Bandelow,  N. Akhmediev, Phys. Lett. A 376 (2012) 1558.  
\bibitem{41}A. Ankiewicz, Y. Wang, S. Wabnitz, N. Akhmediev,  Phys. Rev. E  89 (2014) 012907.
\bibitem{42}Y.V. Bludov, V.V. Konotop, N. Akhmediev, Eur. Phys. J. Spec. Top. 185 (2010) 169. 
\bibitem{43}Z.Y. Yan, V.V. Konotop, N. Akhmediev, Phys. Rev. E 82 (2010) 036610.  
\bibitem{44}Z.Y. Yan, Phys. Lett. A  375 (2011) 4274.
\bibitem{48}Y. Ohta, J.K. Yang, J. Phys. A 46 (2013) 105202. 
\bibitem{49}Zhaqilao, Phys. Lett. A 377 (2013) 855.
\bibitem{50}W.R. Sun, B. Tian, Y. Jiang, H.L. Zhen, Ann. Phys. 343 (2014) 215.
\bibitem{51}F. Baronio,  A. Degasperis, M. Conforti, S. Wabnitz, Phys. Rev. Lett. 109 (2012) 044102. 
\bibitem{52}F. Baronio, M. Conforti, A. Degasperis, S. Lombardo, Phys. Rev. Lett. 111 (2013) 114101. 
\bibitem{53}F. Baronio, M. Conforti, A. Degasperis, S. Lombardo, M. Onorato, S. Wabnitz, Phys. Rev. Lett. 113 (2014) 034101.
\bibitem{lak}N.V. Priya, M. Senthilvelan, M. Lakshmanan,  Phys. Rev. E 88  (2013) 022918.
\bibitem{54}B.L. Guo, L.M. Ling, Chin. Phys. Lett. 28 (2011) 110202. 
\bibitem{55}L.M. Ling, B.L. Guo, L.C. Zhao, Phys. Rev. E 89 (2014) 041201. 
\bibitem{56}L.M. Ling, L.C. Zhao, B.L. Guo, (2014) arXiv:1407.5194.
\bibitem{57}L.M. Ling, L.C. Zhao, B.L. Guo, (2013) arXiv:1309.1037.
\bibitem{58}J.S. He, L.J. Guo, Y.S. Zhang, A. Chabchoub, Proc. R. Soc. A  470 (2014) 20140318.
\bibitem{59}L.C. Zhao,  J. Liu, Phys. Rev. E  87 (2013) 013201.  
\bibitem{60}X. Wang, Y.Q. Li, F. Huang, Y. Chen, Commun. Nonlinear Sci. Numer. Simul. 20 (2015) 434. 
\bibitem{61}R.S. Tasgal, M.J. Potasek, J. Math. Phys. 33 (1992) 1208. 
\bibitem{62}S.H. Chen, L.Y. Song, Phys. Rev. E  87 (2013) 032910.  
\bibitem{63}D. Mihalache, L. Torner, F. Moldoveanu, N.C. Panoiu, N. Truta, Phys. Rev. E 48 (1993) 4699.
\bibitem{64} S.G. Bindu, A. Mahalingam, K. Porsezian, Phys. Lett. A 286 (2001) 321. 
\bibitem{65}K. Porsezian, K. Nakkeeran, Pure Appl. Opt. 6 (1997) L7. 
\bibitem{66}R. Radhakrishnan, M. Lakshmanan, M. Daniel, J. Phys. A 28 (1995) 7299.
\bibitem{67}S.H. Chen, Phys. Lett. A 378 (2014) 2851. 
\bibitem{68}X. Wang, Y.Q. Li, Y. Chen, Wave Motion 51 (2014) 1149.
\bibitem{69}X. Wang, Y. Chen, Chin. Phys. B 23 (2014) 070203.
\bibitem{70}A. Chabchoub, N. Hoffmann, M. Onorato, A. Slunyaev, A. Sergeeva, E. Pelinovsky, N. Akhmediev, Phys. Rev. E 86 (2012) 056601.
\bibitem{71}X. Wang, B. Yang, Y. Chen, Y.Q. Yang, Phys. Scr. 89 (2014) 095210.
\bibitem{72}X. Wang, B. Yang, Y. Chen, Y.Q. Yang, Chin. Phys. Lett. 31 (2014) 090201.
\bibitem{73}C.H. Gu, H.S. Hu,  Z.X. Zhou, Darboux transformations in integrable systems: theory and their applications to geometry, Springer-Verlag, New York, 2005.
\bibitem{74}V.B. Matveev, M.A. Salle, Darboux transformations and solitons, Springer-Verlag, Berlin-Heidelberg, 1991.
\end{thebibliography}
\end{document}